\documentclass[%
 reprint,
 amsmath,amssymb,
 aps, prd,preprintnumbers, superscriptaddress, nofootinbib
]{revtex4-2}

\usepackage{graphicx}	
\usepackage{amsmath}	
\usepackage{amssymb}	

\usepackage[english]{babel}
\usepackage{xspace}
\usepackage{graphicx}
\usepackage{latexsym}
\usepackage{floatflt}
\usepackage{float}
\usepackage{units}
\usepackage{hyperref}
\usepackage{amsmath}
\usepackage{gensymb}
\usepackage{mathdots}
\usepackage{color}
\usepackage{xcolor}
\usepackage{mathtools}
\usepackage{lipsum}

\usepackage{dcolumn}
\usepackage{bm}

\def\aapr{\ref@jnl{A\&A~Rev.}}          
\def\aap{A\&A}
\def\apj{ApJ}
\def\apjs{ApJS}
\def\apjl{ApJL}
\def\mnras{MNRAS}

\def\prd{Phys.Rev.D}         
\def\apss{ApSS}
\def\jcap{JCAP}

\newcommand{\expf}[1]{{{\rm e}^{#1}}}

\newcommand{\TCMB}{T_{\rm CMB}}

\newcommand{\Planck}{{\it Planck}\xspace}

\newcommand{\id}{{\,\rm d}}

\newcommand{\beq}{\begin{equation}}   %

\newcommand{\eeq}{\end{equation}}   %

\newcommand{\beqa}{\begin{eqnarray}}   %

\newcommand{\eeqa}{\end{eqnarray}}   %

\newcommand{\beal}{\begin{align}}

\newcommand{\enal}{\end{align}}

\newcommand{\bspl}{\begin{split}}

\newcommand{\espl}{\end{split}}

\newcommand{\bsub}{\begin{subequations}}

\newcommand{\esub}{\end{subequations}}

\newcommand{\bmulti}{\begin{multline}}   %

\newcommand{\beqm}{\begin{mathletters}}   %

\newcommand{\eeqm}{\end{mathletters}}   %

\newcommand{\me}{m_{\rm e}}

\newcommand{\Ne}{N_{\rm e}}

\newcommand{\pot}[2]{#1 \times 10^{#2}}

\newcommand{\COBEF}{{\it COBE/FIRAS}\xspace}

\newcommand{\PIXIE}{{\it PIXIE}\xspace}
\newcommand{\WMAP}{{\it WMAP}\xspace}

\newcommand{\Tin}{T_{\rm in}}

\newcommand{\zcon}{z_{\rm con}}
\newcommand{\xin}{x_{\rm in}}
\newcommand{\eV}{{\rm eV}}
\newcommand{\keV}{{\rm keV}}

\newcommand{\ma}{m_{\rm a}}

\newcommand{\gammacon}{\gamma_{\rm con}}

\newcommand{\mucon}{\mu_{\rm con}}

\begin{document}

\preprint{MIT-CTP/5723}

\title{Revisiting Constraints on Resonant Axion-Photon Conversions from CMB Spectral Distortions}
\author{Bryce Cyr}
\email{brycecyr@mit.edu}
\affiliation{Center for Theoretical Physics, Massachusetts Institute of Technology, Cambridge, MA 02139,USA}

\author{Jens Chluba}
\email{jens.chluba@manchester.ac.uk}
\author{Pranav Bharadwaj Gangrekalve Manoj}
\email{pranavbharadwaj.gangrekalvemanoj@manchester.ac.uk}
\affiliation{%
Jodrell Bank Centre for Astrophysics, School of Physics and Astronomy, The University of Manchester, Manchester M13 9PL, U.K.
}%

\begin{abstract}
\noindent Axions and axion-like particles (ALPs) remain highly motivated extensions to the standard model due to their ability to address open questions such as the relic abundance of dark matter and the strong CP problem. Axions are also capable of undergoing a resonant mixing with photons when the masses of the two fields are roughly equal, producing a wide array of phenomenological consequences. Here, we revisit constraints coming from conversions of the cosmic microwave background (CMB) into axions, which will induce a distortion to the frequency spectrum of the background photons. We introduce a more detailed description for the modeling of the plasma mass of the photon, showcasing how the inclusion of Helium recombination can alter the conversion probability for photons in the Wien tail. Our results include an updated analytic framework, which allows us to define the precise spectral shape of the axion distortion, as well as a numeric component which utilized the code \texttt{CosmoTherm} to fully characterize the distortion, providing a slight increase in the constraining power over the analytics. We also treat for the first time the large-distortion regime for resonant axion-photon conversions. Under the assumption of large-scale primordial magnetic fields near the limit obtained from CMB observations, we find that spectral distortions can probe previously unexplored regions of the axion parameter space.
\end{abstract}

\maketitle

\section{\label{sec:level1}Introduction}
\noindent To date, the cosmic microwave background (CMB) offers us one of the cleanest and most sensitive probes of the thermal history of the Universe. Details discerned from precise studies of temperature and polarization anisotropies over the past decades by the \WMAP \citep{WMAP7yrCosmo} and \Planck \citep{Planck2018params} satellites have ushered in an age of concordance cosmology in which the bulk features of the Universe can be understood through a handful of parameters in a model known as $\Lambda$CDM. Despite the enormous success of this model, the microphysical properties of dark matter and dark energy remain unknown, and small deviations away from $\Lambda$CDM may still exist. 

While the CMB anisotropies give a well-understood snapshot of the Universe at very early times ($z \simeq 1000$), the frequency spectrum of the CMB is a complementary piece of information which can allow us to probe even further back in time. The most recent space-based measurement of the (sky-averaged) CMB spectrum took place onboard the \COBEF satellite \citep{Fixsen1996,Fixsen2003} nearly three decades ago, revealing an exquisitely precise blackbody shape with temperature $T_{\rm CMB,0} = 2.725 \pm 0.002$ K. This observation constrained departures away from a blackbody at the level of roughly $\Delta I/I \lesssim 10^{-5}$ and to date remains the most perfect blackbody measured in nature. 

The precision of this measurement implies that the pre-recombination plasma was very close to thermal. Thermalization, however, is not an instantaneous process, and requires that interactions which can both redistribute energy across the spectrum (Compton scattering), as well as create and destroy photons (double Compton and Bremsstrahlung) to be efficient. It has been well known for some time that non-thermal injections/extractions of energy or entropy into the pre-recombination bath can produce deviations away from a blackbody \citep{Zeldovich1969,Sunyaev1970,Burigana1991,Hu1993, Chluba2011therm,Sunyaev2013,Tashiro2014}, with distinct spectral shapes classified as $\mu$- and $y$-type distortions. In particular, $\mu$-type spectral distortions can only be generated between $5\times 10^4 \lesssim z \lesssim 2 \times 10^6$, so their detection would be a smoking gun signature of truly primordial physics. 

Though no follow-up measurement of the spectrum has taken place thus far, many proposals exist that could (in principle) allow for an observation of the CMB spectrum at the level of $\Delta I/I \lesssim 10^{-9}$. The most prevalent experimental design is the \PIXIE concept \citep{Kogut2011PIXIE,Kogut2016SPIE,Kogut2024}, which is similar in spirit to \COBEF, and could hit this target by simply leveraging the last three decades of improved detector technology. Other designs are also being considered \citep{PRISM2013WPII, Chluba2021Voyage}, most recently the {\it SPECTER} concept \citep{SPECTER2024} which utilizes three unique detector technologies in different frequency bands. This approach allows one to efficiently mitigate foreground contamination, which is expected to be the limiting factor in sensitivity when searching for the $\mu$-distortion \citep[e.g.,][]{abitbol_pixie, Rotti2021SD}. 

Atmospheric noise makes detection of a very weak $\mu$-distortion untenable from the Earth. Late-time effects present in the standard $\Lambda$CDM cosmology do however provide us with robust sources of $y$-type distortions \citep{Refregier2000,Hill2015,Chluba2016}. Recently, a balloon-borne experiment known as BISOU \citep{BISOU} has moved into Phase A of its development, and is expected to measure the CMB spectrum a factor of $\simeq 20-30$ times more sensitively than \COBEF. This sensitivity should be adequate to detect the largest $y$-distortion sources, meaning that the first ever detection of a CMB spectral distortion is on the horizon. In addition to this, other ground based observations such as COSMO \citep{Masi2021}, TMS \citep{Jose2020TMS}, and APSERa \citep{Mayuri2015} are targeting different frequency bands, and will soon provide us with complementary data also covering the Rayleigh-Jeans tail of the CMB.

Future detections of spectral distortions boast an impressive improvement in constraining power for any beyond standard model (BSM) processes which can disrupt thermal equilibrium. One such model that can be constrained is the presence of axions or axion-like particles (ALPs). These fields are well-motivated pseudo-scalar extensions to the standard model of particle physics capable of transferring energy to and/or from the CMB in the distortion epochs. 

The canonical axion was first proposed as a way to address the strong CP problem in QCD \citep{Peccei1977,Weinberg1977,Wilczek1977}, and culminated in the development of two well-established benchmark models known as the KSVZ \citep{Kim1979,Shifman1979} and DFSZ \citep{Zhitnitsky1980,Dine1981} axions. One characteristic of these so-called QCD axions is that there exists a relationship between the mass ($m_{\rm a}$) and coupling to photons ($g_{\rm a \gamma \gamma}$) that picks out a region of parameter space which is of particular interest to the community (See \citet{Marsh2015} and \citet{OHare2024} for details and updated bounds). 

ALPs, on the other hand, do not present a solution to the strong CP problem, but are nevertheless interesting to study. In these theories, $m_{\rm a}$ and $g_{\rm a \gamma\gamma}$ are independent free parameters which can be constrained based on the myriad of observational phenomena they can induce. UV completions to the standard model also seem to indicate that a plethora of ALP fields may be produced at very early times, in a scenario known as the string axiverse \citep{Arvanitaki2009}.

In the presence of magnetic fields, resonant conversions can take place between photons and axions/ALPs through a Chern-Simons type interaction, as originally discussed by \citet{Yanagida1987}. The direction of the conversion is dependent on the relative occupation between the two sectors, a fact that can allow distortions to be sourced either in the presence of an abundant axion sector ($\Omega_{\rm a} \simeq \Omega_{\rm CDM}$), or an initially unpopulated state ($\Omega_{\rm a} \ll 1$). One can also consider constraints on axions coming from perturbative decays. The benefit of this is that magnetic fields are no longer necessary, though the constraints from spectral distortions \citep{Bolliet2020dp} and the Lyman-alpha forest \citep{Capozzi2023} from this process are superseded by more precise experiments. In this work we focus on resonant conversions in the $\Omega_{\rm a} \ll 1$ scenario.

Here, we build off of previous results which considered spectral distortions sourced by resonant conversions of axions/ALPs into photons \citep{mirizzi2009constraining, tashiro2013constraints,Mukherjee2018} in the presence of a primordial magnetic field. Conversions of this type induce a deficit in CMB photons, which will show up as $\mu$- or $y$-type distortions, as we illustrate here. Using an updated analytic formulation, as well as state-of-the-art numerical computations with {\tt CosmoTherm} \citep{Chluba2011therm}, we derive constraints based on a re-analysis of the \COBEF residuals over a wide range of axion masses. This work considers global (sky-averaged) spectral distortions, making it complementary to recent results discussing axion constraints from the patchy screening mechanism \citep{Mondino2024, goldstein2024constraints}, which takes advantage of the variation of magnetic fields and plasma densities within large scale structure.

We improve upon other works in a number of ways. First, we refine the modeling of the effective photon mass by more consistently including the absorpative effects of neutral hydrogen after recombination, while also including contributions from helium. Second, we more consistently include the effects of entropy (direct photon) extraction on the distortion signature by performing a reanalysis of the \COBEF data in the presence of late-time conversions. Third, we discuss the large distortion regime in which a significant fraction of the CMB can convert into axions ($\Delta \rho/\rho \lesssim 1$). What is presented here builds directly on our recent work on photon to dark photon conversion processes \citep{Chluba2024DP}, where a significant amount of the required machinery was developed.

The remainder of the paper is organized as follows: in Sec.~\ref{sec:level2}, we discuss axion-photon couplings in more detail, highlighting our procedure to model the photon thermal mass. Next, we discuss the generation of spectral distortions from this mechanism in Sec.~\ref{sec:level3}. In that section, we discuss our formalism for both the perturbative limit, as well as the large distortion regime. Section~\ref{sec:level4} is dedicated to understanding the precise form of solutions to the photon Boltzmann equation. Finally, we quote updated limits on the axion-photon parameter space in Sec.~\ref{sec:level5} before concluding in Sec.~\ref{sec:level6}. Here we work in natural units where $\hbar = k_{\rm b} = c = 1$, and use the terms axion and ALP interchangeably to mean a pseudoscalar extension to the standard model. 
\section{Axion-Photon Couplings}
\label{sec:level2}
\noindent At lowest order, the axion can couple to photons through a Chern-Simons interaction of the form
\begin{align} \label{eq:int-term}
    \mathcal{L}_{\rm int} = -\frac{g_{\rm a \gamma \gamma}}{4} \, {\rm a} F_{\mu \nu} \Tilde{F}^{\mu\nu} = g_{\rm a \gamma\gamma} \, {\rm a} \, \textbf{B}\cdot \textbf{E}.
\end{align}
Here, {\rm a} is the axion field, $F_{\mu\nu}$ is the electromagnetic field strength tensor, and $\Tilde{F}^{\mu\nu} = \epsilon^{\mu\nu\alpha\beta} F_{\alpha\beta}/2$ is its dual. The coupling $g_{\rm a \gamma\gamma}$ is a dimensionful quantity whose amplitude is typically related to the axion decay constant $f_{\rm a}$ through $g_{\rm a \gamma\gamma} \propto 1/f_{\rm a}$. The dot product encompasses the fact that for a constant transverse magnetic field, only photons polarized parallel to the $B$-field will convert into axions \citep{Raffelt1987, Mirizzi2005SN}.

\subsection{\label{sec:level3-2}Resonant Conversions}
\noindent In addition to the perturbative decay, axions and photons can mix through a resonant two level system, reminiscent of the Mikheev-Smirnov-Wolfenstein (MSW) effect which describes neutrino oscillations in matter. For photon-axion conversions, the presence of a background magnetic field is a necessary ingredient to mediate the conversion. Here, we will consider only the conversion of CMB photons into axions in the scenario where the initial axion sector is unpopulated ($\Omega_{\rm a} \simeq 0$). A resonant conversion takes place when the matching condition $m_{\rm a} \simeq m_{\gamma}$ is satisfied, where $m_{\gamma}$ is the mass of the photon induced by thermal interactions in the plasma. We discuss this quantity in detail in the following subsection.

The two level system describing conversions from photons to axions is most often solved using the well-known Landau-Zener formalism, whose details (as well as caveats) have been explored recently in a number of papers \citep{Brahma2023,Carenza2023, Brahma2024}. For application to spectral distortions, we follow the formalism laid out in \citet{Chluba2024DP} where a similar problem was explored for photon to dark photon conversions. In this setup, it is useful to parameterize the conversion probability as 
\begin{align}
\label{eq:Pcon_simp}
    P_{\rm \gamma_{||} \rightarrow a}(\gammacon, x) \simeq 1-\exp\left(-\gammacon x\right),
\end{align}
where we use the dimensionless (and redshift independent) frequency $x=\omega/\TCMB$ and define the conversion parameter
\begin{subequations}
\begin{align} \label{eq:gamma_con_etc}
\gammacon
&=
\left.\frac{\pi \,\kappa^2 \TCMB(z) (1+z)^4}{\ma^2 H(z)\, (1+z)} \left|\frac{\id \ln m^2_{\gamma}}{\id z} \right|^{-1}\right|_{z=\zcon},
\\[1mm]
\kappa &= g_{\rm a \gamma \gamma} B^0_{\rm rms, T}\approx\pot{1.95}{-30}\,{\rm eV}\,\epsilon,
\\[1mm]
\label{eq:def_eps}
\epsilon&=\left[\frac{g_{\rm a \gamma \gamma} B^0_{\rm rms, T}}{10^{-10}\,{\rm GeV}^{-1}\,{\rm nG}}\right],
\end{align}
\end{subequations}
where $T_{\rm CMB}(z)$ is the microwave background temperature, $H(z)$ is the Hubble rate, and $B^0_{\rm rms} = \sqrt{\langle B^2 \rangle}$ is the root mean squared value of the comoving magnetic field strength, whose transverse component relative to the direction of a propagating photon ($B_{\rm rms, T}^0$) can be calculated given a specific scenario. An upper bound on the amplitude of the primordial magnetic field today can be derived using the \Planck dataset \citep{Planck2018params}, and is given roughly by $B^0_{\rm rms} \lesssim 1 \, {\rm nG}$ for quasi-scale invariant primordial magnetic fields (PMFs)\citep{Ade:2015cva, Paoletti2022}, although one should note that the constraints strongly depend on the assumed production mechanism \citep[e.g.,][for PMF physics and perspectives]{Subramanian1998, Banerjee2004, Durrer2013, Ade:2015cva, Jedamzik2018}. We have also implicitly assumed in the expression for $\gamma_{\rm con}$ that the amplitude of the $B-$field scales as $B_{\rm rms,T}(z) = B_{\rm rms,T}^0(1+z)^2$, as appropriate for a conserved comoving amplitude. This assumption is not crucial when considering considering axion masses $\ma > \pot{\rm few}{-10}\,\eV$; however, our constraints cannot be easily mapped for masses $\ma \lesssim \pot{\rm few}{-10}\,\eV$, since for these fields multiple conversions can occur. In the latter regime, it is possible to analytically model the distortion for specific models of $B_{\rm rms,T}(z)$, as we show below.

In contrast to the dark photon case, high frequency photons convert preferentially into axions, and under certain conditions, large deficits can appear in the Wien tail of the CMB during a conversion process. Also in contrast is the fact that only photons polarized parallel ($\gamma_{||}$) to $B^0_{\rm rms, T}$ will be converted into axions, leading to both intensity and polarization fluctuations in the CMB \citep{Mukherjee2018,goldstein2024constraints}. For highly non-adiabatic ($\gamma_{\rm con} \ll 1$) conversions of the CMB monopole, the usual Landau-Zener expression holds \citep{Carenza2023}, although it may be necessary to apply a more careful approach in the large distortion limit, where $\gamma_{\rm con} \gtrsim 1$.

For the Landau-Zener formalism to be applicable, the coherence length of the $B$-field should be greater than the typical oscillation length scale of photons into axions, in other words $\ell_{\rm osc} \lesssim \ell_{\rm B}$\footnote{In principle a more complete hierarchy is $\ell_{\rm osc} \lesssim \ell_{\rm pl} \lesssim \ell_{\rm B}$ where $\ell_{\rm pl}$ is the typical length scale of homogeneity in the plasma. For the generation of CMB spectral distortions, this turns out to be satisfied for situations with small $\gamma_{\rm con} $\citep{Carenza2023}}. It can be shown \citep{Raffelt1987,Marsh2021,Mondino2024} that for magnetic coherence lengths satisfying
\begin{align} \label{eq:B-field-coh}
    \ell_{\rm B} \gg \frac{4\pi \omega}{m_{\rm a}^2} &\simeq 10^{-2} \, {\rm pc} \left( \frac{\omega}{10^{-4} \, {\rm eV}}\right) \left( \frac{10^{-12} \, {\rm eV}}{m_{\rm a}}\right)^2 \nonumber\\
    &\simeq 66 \, {\rm pc} \left(\frac{x}{2.82} \right)\left(\frac{1+z}{10^3} \right) \left( \frac{10^{-12} \, {\rm eV}}{m_{\rm a}}\right)^2
\end{align}
the conversion probability is accurately represented by Eq.~\eqref{eq:Pcon_simp}. Note here that $x = 2.82$ is the redshift-independent value for the peak frequency of the CMB. Upper bounds from \Planck are given on scales of roughly $\simeq 1 \, {\rm Mpc}$, so for the range of axion masses considered in this work, the condition posed in Eq.~\eqref{eq:B-field-coh} is always satisfied. 


For situations involving multiple resonant conversions, it becomes more complicated to solve the general time-dependent radiative transfer equation for the CMB. Nevertheless, one can generalize the conversion coefficient to
\begin{align}
\gammacon\approx
\sum_i\left.\frac{\pi \,\kappa^2 \TCMB(z) (1+z)^4}{\ma^2 H(z)\, (1+z)} \left| \frac{\id \ln \omega^2_{\gamma}}{\id z} \right|^{-1}\right|_{z=z_{{\rm con}, i}}
\end{align}
when computing the conversion probability in Eq.~\eqref{eq:Pcon_simp}. Here the sum is over the redshifts at which the resonance condition is fulfilled. The main effects can all be captured analytically, as we explain below. 

Allowing for multiple level crossings is known to come with some additional subtleties. One such possibility is that of phase interference in which the amplitude of the conversion can be damped when level crossings occur in quick succession. For monopole conversion processes in the cosmological context, \citet{Brahma2023} have shown that phase interference effects can be neglected when
\begin{align} \label{eq:Brahma-MLC}
    \left| \int_{z_1}^{z_2} \frac{\id z}{(1+z)H(z) }\frac{m_{\rm a}^2 - m_{\gamma}^2(\omega,z)}{2\omega} \right| \gg 2\pi,
\end{align}
where $z_1$ an $z_2$ are the redshifts at which the resonant condition $m_{\rm a} \simeq m_{\gamma}$ are met for a photon with frequency $\omega$. As we show in the following subsection, even in the simple monopole case the structure of the photon thermal mass $m_{\gamma}$ can be complicated. For low frequency photons, $m_{\gamma}(\omega,z) \simeq m_{\gamma}(z)$, and the thermal mass decreases monotonically (neglecting for the moment reionization). This implies that Rayleigh-Jeans photons only ever encounter one level crossing, while higher energy photons can encounter multiple level crossings even in the monopole case after recombination. For a given axion mass, there will typically be an extremely tight frequency range in which Eq.~\eqref{eq:Brahma-MLC} is not satisfied between multiple crossings. However, the relative energy density stored in this tight bandwidth during a resonant conversion is typically negligble, and so we neglect the details of this effect in what follows.

\subsection{Conversion redshifts at high frequencies}
\noindent Given an axion mass, the first step is to determine the conversion redshifts when $m_{\rm a} \simeq m_{\gamma}$. We start from the general expression for an effective photon mass in plasma \citep[e.g.,][]{Hecht2017}
\begin{align}
\label{eq:mgamma_def}
m^2_\gamma&=-\omega^2 ({\rm n}^2-1),
\end{align}
where ${\rm n}$ denotes the refractive index of the medium, which generally is (strongly) frequency dependent. Consider an unmagnetized\footnote{Although below we will assume the presence of a large-scale magnetic, the related corrections to the dispersion relation are only important at very low frequencies well below the Bremsstrahlung and double Compton cut-offs (i.e., outside of the thermalization domain). At larger magnetic field strengths, the corrections should possibly be considered more carefully; however, this problem is beyond the scope of this work.} plasma with free electrons, neutral hydrogen, helium and singly-ionized helium atoms. In this background, the refractive index behaves as
\begin{align}
\label{eq:n2m1_def}
{\rm n}^2-1=({\rm n}^2&-1)\big|_{\rm e} +({\rm n}^2-1)\big|_{\rm HI} \nonumber \\
&+({\rm n}^2-1)\big|_{\rm HeI}+({\rm n}^2-1)\big|_{\rm HeII}.
\end{align}
The electron contribution is simply given by 
\begin{align}
({\rm n}^2-1)\big|_{\rm e}&=\frac{N_{\rm e} e^2}{\me}\frac{1}{-\omega^2+i \gamma_{\rm e} \omega}=-\frac{\omega_{\rm p}^2}{\omega^2}\frac{1}{1-i (\gamma_{\rm e}/\omega)},
\end{align}
where $N_{\rm e}$ is the free electron number density, $m_{\rm e}$ the electron mass, and the electric charge is $e^2 \simeq 4\pi/137$ in natural units. Note also that $\gamma_{\rm e}$ is the (dimensionful) damping coefficient for free electrons. At sufficiently high frequencies, $\gamma_{\rm e}$ is usually neglected, although at low frequencies one expects corrections from collisions, free-free absorption and double Compton scattering which we omit here. We have also introduced the plasma frequency 
\begin{align}
\omega_{\rm p}=\sqrt{\frac{N_{\rm e} e^2}{\me}}=\tilde{\omega}_{\rm p}\,\sqrt{X_{\rm e}},
\end{align}
where $X_{\rm e} = N_{\rm e}/N_{\rm H}$ is the free electron fraction. If no other contributions to the refractive index exist, with Eq.~\eqref{eq:mgamma_def} this implies
$$m^2_\gamma\approx \omega_{\rm p}^2.$$ This condition is commonly used for the pre-recombination plasma. This expression implies single resonant conversions at a given mass as long as $\ma\gtrsim 10^{-12}\,\eV$ even when the average reionization history is included (see below).
 
 To add the contributions from the atomic species, we can simply think of the atoms as a collection of oscillators. Neutral hydrogen atoms are primarily in the ground-state, meaning that only contributions involving the Lyman-series transition or transitions to the continuum really contribute. All of these will be excited in some cases by virtual photons. 
Only considering the Lyman-series contributions, we have \citep{Hecht2017}
\begin{align}
\label{eq:n2m1_def_H}
({\rm n}^2-1)\big|_{\rm HI}&=\frac{N_{\rm HI} e^2}{ \me}\sum_{j=2}^\infty \frac{f_j}{\omega_j^2-\omega^2+i \gamma_j \omega},
\end{align}
where $N_{\rm HI}$ is the fraction of neutral hydrogen, and $\omega_j=\omega_R(1-j^{-2})$ is the Lyman series transition frequency  with $\omega_R\approx 13.6\,\eV$ and $\gamma_j$ the corresponding damping coefficient\footnote{These can usually be estimated using the lifetime of the excited state, i.e., $\gamma_j \propto A_{j\rightarrow {\rm 1s}}$, where $A_{j\rightarrow {\rm 1s}}$ is the transition rate to the ground state. However, we will not consider the regions close to the resonances in this work.}. The oscillator strength of the Lyman transition is \citep{Rybicki1979}
$$f_j=\frac{2^8}{3} j^5 \frac{(j-1)^{2j-4}}{(j+1)^{2j+4}}.$$
Using $N_{\rm HI}=N_{\rm H} (1-X_{\rm p})$, we can write $\frac{N_{\rm HI} e^2}{ \me}\equiv \tilde{\omega}_{\rm p}^2 (1-X_{\rm p})$, implying 
\begin{align}
({\rm n}^2-1)\big|_{\rm HI}&=\tilde{\omega}_{\rm p}^2 (1-X_{\rm p}) \kappa_{\rm HI}(\omega).
\end{align}
Here we introduced $\kappa_{\rm HI}(\omega)$ to denote the sum in Eq.~\eqref{eq:n2m1_def_H}. This quantity is generally strongly frequency dependent. In particular close to the Lyman-series transition frequencies, significant damping contributions appear; however, we will neglect these complications here. 

Assuming $\omega \ll \omega_2$ (redward of the Lyman-$\alpha$ resonance), we find the approximation
$$\kappa_{\rm HI}\approx \sum_{j=2}^\infty \frac{f_j}{\omega_j^2}=\frac{1}{\omega_R^2} \sum_{j=2}^\infty \frac{2^8}{3} j^9 \frac{(j-1)^{2j-6}}{(j+1)^{2j+6}} \approx \frac{0.0050}{{\rm eV}^2}.$$ This then yields
\begin{align}
\label{eq:mgamma_def_eH_old}
m^2_\gamma&=\tilde{\omega}_{\rm p}^2\left[X_{\rm e}- \pot{5.0}{-3}\left(\frac{\omega}{{\rm eV}}\right)^2\,X_{\rm HI}\right],
\end{align}
where we have included the Lyman-series transitions up to $j=2000$. Here, $X_{\rm HI}=N_{\rm HI}/N_{\rm H}\equiv (1-X_{\rm p})$ is the neutral hydrogen fraction.

The coefficient for the contribution from hydrogen differs from the quantity sometimes seen in the literature, i.e. $\pot{7.7}{-3}$ \citep{Wondrous2020} or $7.3\times 10^{-3}$ \citep{Mukherjee2018}. However, here it should be noted that these values appear to be derived based on the refractive index measurements of {\it molecular} hydrogen, which is not directly applicable. At 1 atm and an ambient temperature of $273$~K, atomic hydrogen cannot exist in significant amounts unless {\it in statu nascendi}, which renders a laboratory measurement challenging. In the universe, the fraction of ${\rm H}_2$ is extremely small until the reionization era is reached \citep{Stancil1996, Stancil1998}. We note that our approach finds relative agreement with \citet{Berlin2022} who computed an $X_{\rm HI}$ coefficient of $\simeq 5.3\times 10^{-3}$ using a fluid-based approach.

In principle, it is still possible to estimate the hydrogen contribution to the cosmic photon dispersion relation from lab measurements of ${\rm H}_2$ after a few small modifications.
From the literature values one has \citep{Hecht2017} $({\rm n}-1)\big|_{\rm H_2}\approx \pot{1.32}{-4}$ or $({\rm n}^2-1)\big|_{{\rm H}_2}\approx \pot{2.62}{-4}$ at standard temperature and pressure (STP). With a number density of molecules of $N_{{\rm H}_2}=P/kT \approx \pot{2.68}{19} {\rm cm^{-3}}$ we then obtain 
\begin{align}
\label{eq:kappa_H2}
\kappa_{{\rm H}_2}\approx \frac{({\rm n}^2-1)\big|_{{\rm H}_2}}{N_{{\rm H}_2}\,\frac{e^2}{\me}}\approx \frac{\pot{7.10}{-3}}{{\rm eV}^2},
\end{align}
which is very close to the one often used. Redoing the computation with the measured density of ${\rm H}_2$ under normal conditions (this would include van der Waals corrections etc.), we have $N_{{\rm H}_2}\approx \pot{2.50}{19} {\rm cm^{-3}}$ and hence $\kappa_{{\rm H}_2}\approx \frac{\pot{7.6}{-3}}{{\rm eV}^2}$, which is again very close to the quoted value. However, this is still {\it not} the effective value that would apply for atomic hydrogen, as we also have to divide by a factor of two, given that ${\rm H}_2$ has two electrons per molecule. The resulting effective coefficient is then $\kappa_{{\rm HI}}\approx \frac{\pot{3.8}{-3}}{{\rm eV}^2}$, which is about $30\%$ lower than the theoretical value we estimated above. 

\begin{figure*}
\includegraphics[width=1.8\columnwidth]{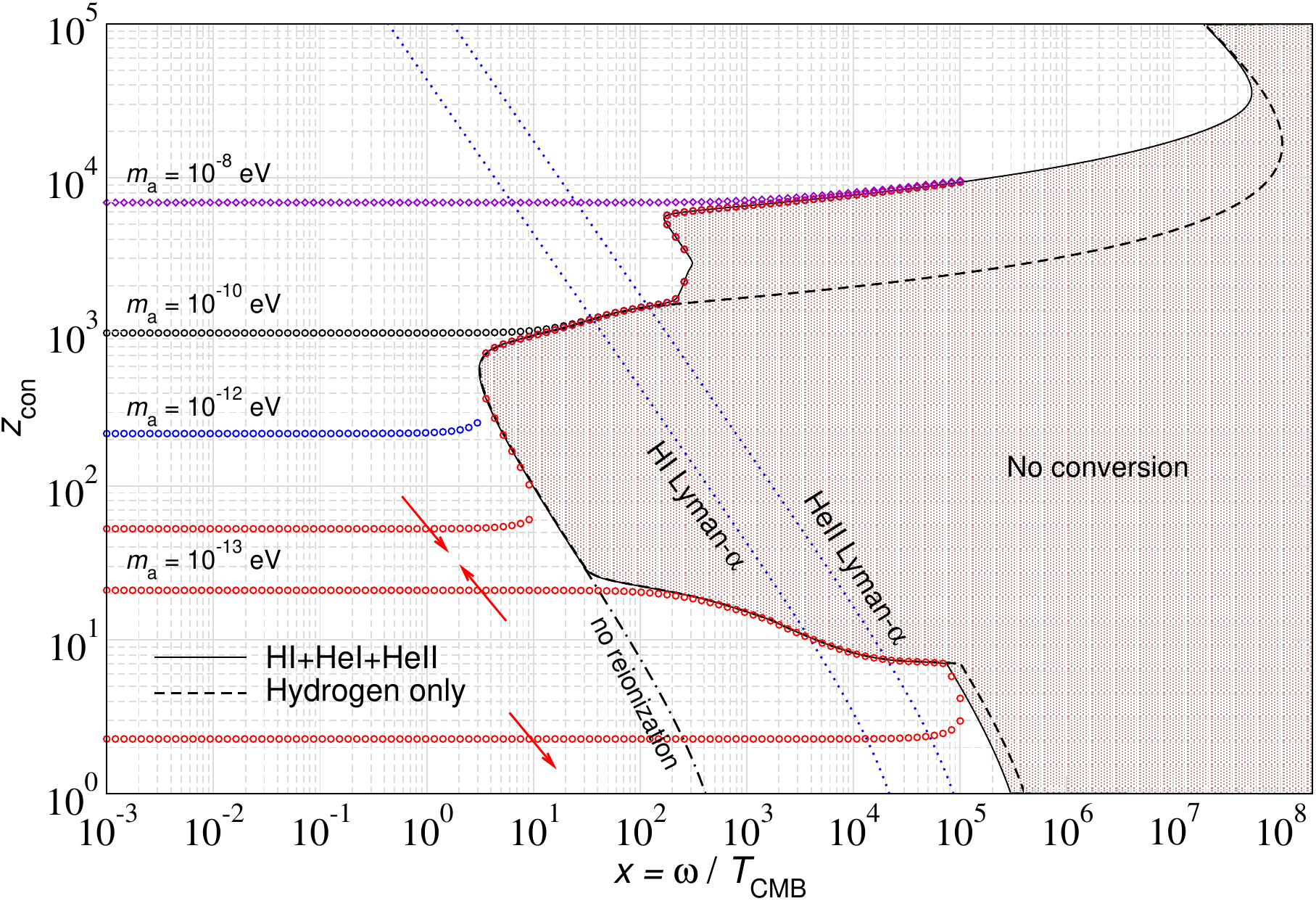}
\caption{Conversion redshifts for several axion masses. We solved the condition $\ma=m_\gamma$ numerically using the standard recombination history from {\tt CosmoRec}. We illustrate the modifications to the domains in which no conversions occur when only including hydrogen. The blue dotted lines are the loci of the HI and HeII Lyman-$\alpha$ lines for reference. Neglecting reionization further changes the no conversion domain at low redshifts. The red arrows indicate in which direction the conversion loci move when further lowering the axion mass.}.
\label{fig:x_gamma_crit}
\end{figure*}

For the contributions of HeI and HeII, a very similar procedure applies, yielding the more complete expression
\begin{align}
\label{eq:mgamma_def_eH}
m^2_\gamma&\approx \tilde{\omega}_{\rm p}^2\left[X_{\rm e}-\frac{\omega^2}{{\rm eV}^2}\, \sum_{\rm i} \kappa_{\rm i} X_{\rm i}\right],
\end{align}
where ${\rm i} = {\rm e, HI, HeI, HeII}$. Since the number fraction of helium nuclei is $\simeq 8\%$, the contributions from HeI and HeII remain much smaller. In addition, since $\omega_j^{\rm HeII}\approx 4 \omega_j^{\rm HI}$, we have $\kappa_{\rm HeII}\approx \kappa_{\rm HI}/16\approx \pot{3.1}{-4}\,{\rm eV}^{-2}$.

For neutral helium, we include all singlet-singlet transitions from the ground state using the helium atom that was developed for computations of the cosmological recombination history \citep{Chluba2010b, Chluba2012HeRec, Chluba2016CosmoSpec} and include the multiplet tables from \citet{Drake2007}. This yields $\kappa_{\rm HeI}\approx \pot{1.8}{-3}\,{\rm eV}^{-2}$ and includes a factor of two because there are two electrons per neutral helium atom. Using the measured literature value $({\rm n}-1)\big|_{\rm HeI}\approx \pot{0.36}{-4}$ (HeI is indeed a monoatomic gas under normal conditions), we have $({\rm n}^2-1)\big|_{{\rm HeI}}\approx \pot{7.2}{-4}$ and hence $\kappa_{\rm HeI}\approx \pot{2.1}{-3}\,{\rm eV}^{-2}$. Comparing to the naive expectation of $\kappa_{\rm HeI}\approx 2\,\kappa_{\rm HI}/(2.08)^2\approx \pot{2.3}{-3}\,{\rm eV}^{-2}$ from the simple fact that $\omega_2^{\rm HeII}\approx 2.08\,\omega_2^{\rm HI}$ also shows reasonable agreement. 

It is beyond the scope of this work to compute more accurate relations for the effective photon mass. In our computations, we did not include the effects from transition to the continuum nor do we have the means to estimate collisional corrections. In the discussion below, we will use the expression in Eq.~\eqref{eq:mgamma_def_eH} with $\kappa_{\rm HI}\approx \pot{5.0}{-3}\,{\rm eV}^{-2}$, $\kappa_{\rm HeI}\approx
\pot{1.8}{-3}\,{\rm eV}^{-2}$ and $\kappa_{\rm HeII}\approx \pot{3.1}{-4}\,{\rm eV}^{-2}$. Occasionally, we omit the contributions from helium to illustrate its effect. We note, however, that the resonance structure of the refractive index is expected to lead to multiple conversions even in the pre-recombination era, which we plan to investigate in the future. 

\subsubsection{Determination of the conversion redshifts}
\noindent Given the expression for the effective photon mass, we can numerically determine the conversion redshifts ($z_{\rm con}$) from the condition $\ma^2 \approx m_\gamma^2$. Since the terms $\propto \omega^2$ can dominate over the plasma contribution, it is first useful to determine critical frequency
\begin{align}
\label{eq:omega_f_def_eH}
\omega_{\rm f}&= 
\left[\frac{X_{\rm e}}{\kappa_{\rm HI}\,X_{\rm HI}+\kappa_{\rm HeI}\,X_{\rm HeI}+\kappa_{\rm HeII}\,X_{\rm HeII}}\right]^{1/2}\,{\rm eV}
\end{align}
above which there is certainly no conversion, as $m_{\gamma}^2 < 0$. In Fig.~\ref{fig:x_gamma_crit}, we illustrate the shape of this domain. We use {\tt CosmoRec} \citep{Chluba2010b} to compute the hydrogen and helium fractions. Instead of $\omega_{\rm f}(z)$ we show $x_{\rm f}=\omega_{\rm f}(z)/\TCMB(z)$. At $x\lesssim 3.1$, the low frequency approximation ($\omega\rightarrow 0$) for the effective photon mass is fairly accurate and implies single conversions for most axion masses.\footnote{The value often quoted in the literature is $x\lesssim 3.16$; however, this assumes $\kappa_{\rm HI}\approx \pot{7.7}{-3}\,{\rm eV}^{-2}$ based on erroneous arguments.}
At higher frequencies, significant changes in the domains for which conversion is prohibited appear when omitting the contributions from helium or the reionization process. However, this mainly affects frequencies\footnote{Only about $\Delta \rho/\rho \simeq 10^{-6}$ of the CMB energy density is present at $x>20$.} $x>20$, which are not of great importance when it comes to the conversion of CMB photons.

In Fig.~\ref{fig:x_gamma_crit}, we also illustrate the loci of conversions for a few examples. We determined these numerically by spanning the redshift domain $1\leq z \leq 10^4$ with a logarithmic grid ($\delta z/z \simeq 10^{-2}$) and then determining the rough locations of conversions on this grid. Once regions of interest were identified, we performed a bisection search to locate the precise values for the conversion redshifts.

Let us first consider the domain $x\lesssim 3.1$. For $\ma\gtrsim 10^{-12}\,\eV$, we only find single conversion redshifts for the standard recombination history. This assumes that the conversion process does not alter the recombination history, which would otherwise be constrained by CMB anisotropy data. However, at $\ma\lesssim 10^{-12}\,\eV$, multiple conversion can occur due to the presence of reionization.\footnote{For the reionization history we use, we find multiple conversions at low frequencies start at axion masses $\ma\lesssim \pot{4.2}{-13}\,\eV$}. This is illustrated for $\ma\lesssim 10^{-13}\,\eV$, which converts at three instances. The direction of the red arrows in this figure indicate how the conversion loci will evolve as the mass is lowered below $m_{\rm a} = 10^{-13}$ eV.

In the domain $x\gtrsim 3.1$, the situation can become more complicated, as multiple conversions can occur close to $x\simeq x_{\rm f}(\zcon)$. Assuming that we do not care about conversions at $x>20$, this implies that for $\ma\lesssim \pot{\rm few}{-10}\,\eV$ one may have a more complicated conversion structure at high frequencies. We will study these scenarios more carefully in Sect.~\ref{sec:level4}. However, at $\ma\gtrsim \pot{\rm few}{-10}\,\eV$, we do not expect significant modifications. We also note that for $\ma\lesssim 10^{-14}\,\eV$ no conversions occur without the inclusion of refractive effects from hydrogen and helium. Indeed for these masses, all conversions occur very close to $x_{\rm f}$.

\section{Axion-Induced CMB Spectral Distortions}
\label{sec:level3}
\noindent To obtain an understanding of the axion distortion, we closely follow \citet{Chluba2024DP}. We start by discussing the regime in which a small fraction of energy/entropy is extracted from the CMB, and then discuss the changes in the large distortion limit. During a resonant conversion, the extraction of CMB photons will induce a sharp (energy-density based) temperature jump in the background at the corresponding redshift. For the case of a single conversion, we define $T_{\rm in}$ as the initial temperature, which drops to $T_{\rm CMB}$ after conversion. Here we focus on the differences in the calculations, and refer the interested reader to \citet{Chluba2024DP} for details and other numerical aspects. 

\subsection{Small distortion evolution}

\subsubsection{$\mu$-era}
\noindent Assuming a single conversion of photons into dark photons, the final $\mu$-parameter can be reliably estimated by \citep{Chluba2014, Chluba2015GreensII}
\begin{align} 
\label{eq:mu-GF}
    \mucon \simeq \frac{3}{\kappa} \,\left[\frac{\Delta \rho_\gamma}{\rho_\gamma}\Bigg|_{\rm con}  - \frac{4}{3} \frac{\Delta N_\gamma}{N_\gamma}\Bigg|_{\rm con}\right] J_{\mu}(\zcon),
\end{align}
where $\kappa \simeq 2.1419$ [i.e., $3/\kappa \approx 1.4007$] is a dimensionless numerical factor, $\epsilon_\rho=\Delta \rho_\gamma/\rho_\gamma\big|_{\rm con}$ and $\epsilon_N=\Delta N_\gamma/N_\gamma\big|_{\rm con}$ respectively denote the fractional change of the photon energy and number density with respect to the {\it initial} blackbody\footnote{These should not be confused with the coupling strength $\epsilon$, which has {\it no} subscript throughout this paper.}. $J_{\mu}(z)$ is the $\mu$-distortion visibility function, which we approximate by
\begin{align}
    J_{\mu}(z) \approx J_{\rm bb}(z) \left[ 1 - {\rm exp}\left( - \left[\frac{1+z}{5.8 \times 10^4}\right]^{1.88}\right)\right]
\end{align}
and $J_{\rm bb}(z) \approx {\rm e}^{-(z/z_\mu)^{5/2}}$ where $z_\mu\approx \pot{1.98}{6}$ marks the freeze out redshift of double Compton scattering. This simply picks out the energy/entropy changes during the $\mu$-distortion epoch.

The small distortion regime ($\epsilon_{\rho} \ll 1$) necessarily implies that $\gammacon\ll 1$. This means that right after the conversion we can write the distortion with respect to the {\it initial} blackbody at the temperature $\Tin>\TCMB$ as 
\begin{align} 
    \Delta n_{\rm in} &\equiv n_{\rm bb}\left(\xin\right)\,\exp\left(- \frac{\gammacon \xin\Tin}{\TCMB}\right)-n_{\rm bb}\left(\xin\right) \nonumber \\  
    &\approx - \gammacon \xin \,n_{\rm bb}(\xin) \nonumber.
\end{align}
The last expression comes from taking the $\xin=\omega / \Tin< \gammacon^{-1}$ limit. In the small distortion regime, $\Tin/\TCMB-1\ll 1$ allowing us to approximate $\TCMB/\Tin\approx 1$ in the conversion probability.
From this initial form of the distortion, we can compute the fractional change of energy and entropy in terms of numerical $G_{\rm k}$ factors, given by  $G_k=\int \frac{x^k}{{\rm e}^x-1} {\rm d} x$. Explicitly, $G_2\approx 2.4041$, $G_3\approx 6.4939$, and $G_4\approx 24.886$, as needed below. One finds
\begin{subequations}
\begin{align}
\epsilon_\rho&=
\frac{\Delta \rho_\gamma}{\rho_\gamma}\Bigg|_{\rm con}
\approx-\gammacon\, \frac{G_4}{G_3}
\approx - 3.8322 \,\gammacon
\\
\epsilon_N&=
\frac{\Delta N_\gamma}{N_\gamma}\Bigg|_{\rm con}
\approx -\gammacon\, \frac{G_3}{G_2}
\approx - 2.7012 \, \gammacon. 
\label{eq:frac_dN}
\end{align}
\end{subequations}
Putting things together, we then obtain
\begin{align} \label{eq:mu-approx}
\mucon\approx -0.3231\,\gammacon\,J_\mu(\zcon).
\end{align}
In contrast to the dark photon scenario \citep{Chluba2024DP}, the conversion of photons happens primarily at high frequencies, thus the contribution from $\epsilon_N$ is less important and cannot overcome the energy removal term. By ignoring the entropy extraction, we would have naively obtained $\mucon\approx-5.3678\,\gammacon \,J_\mu(\zcon)$, a factor of about $17$ times stronger than the true signal as indicated in Eq.~\eqref{eq:mu-approx}.

\vspace{-4mm}
\subsubsection{$y$-era and $\mu/y$-transition}
\noindent As explained in \citet{Chluba2024DP}, the overall effective energy release can be used to model constraints in the $y$-era as well as in the transition to the $\mu$-era. This is simply given by
\begin{align}
\label{eq:limit_total}
\frac{\Delta \rho_\gamma}{\rho_\gamma}\Bigg|_{\rm dis}&\approx \left[\frac{\Delta \rho_\gamma}{\rho_\gamma}\Bigg|_{\rm con}-\frac{4}{3}\,\frac{\Delta N_\gamma}{N_\gamma}\Bigg|_{\rm con}\right]\,J_{\rm bb}(\zcon)
\nonumber\\[1.5mm]
&\approx-0.2307\,\gammacon \,J_{\rm bb}(\zcon),
\end{align}
where we used the approximations from above.
Given an axion model, we can obtain $\Delta \rho_\gamma/\rho_\gamma\big|_{\rm dis}$ and compare this with the \COBEF (i.e., $\Delta \rho_\gamma/\rho_\gamma\big|_{\rm dis}\lesssim \pot{6}{-5}$ at 95\% c.l.) and \PIXIE limits on $\Delta \rho_\gamma/\rho_\gamma$. Indeed this approach yields consistent constraints for {\it all} axion masses below $\lesssim 10^{-4}\,\eV$ (see Sect.~\ref{sec:level5}).

For axion masses $\ma>10^{-8}\,\eV$, one can assume $m^2_\gamma\approx \omega_{\rm p}^2 \propto \Ne(z) \propto (1+z)^3$. This implies $\id \ln m^2_\gamma/\id z\approx 3/(1+z)$. We furthermore have $(1+\zcon) \approx \pot{1.49}{5}\,[\ma/10^{-6}\,\eV]^{2/3}$, which implies $\gammacon\propto (1+\zcon)^5 /[H(\zcon) \,\ma^2]\approx (1+\zcon)^3/\ma^2 \propto {\rm const}$. Given that $\Delta \rho_\gamma/\rho_\gamma\big|_{\rm dis}\propto \gammacon$, we therefore expect the spectral distortion constraints to be mostly independent of mass in the $\mu$ or $y$-distortion eras \citep[see also][]{mirizzi2009constraining}.

\subsection{Large distortion regime}
\label{sec:level3-1}

\subsubsection{Initial spectrum}
\noindent For the large distortion regime we have to more carefully consider the initial condition \citep{Chluba2024DP}, as it is possible for the background temperature to exhibit large jumps during a conversion process. One has the following relationships \citep{Chluba2020large, Acharya2022large}
\begin{align}
\frac{\Delta\Tin}{\TCMB}&=-\frac{(1+\epsilon_\rho)^{1/4}-1}{(1+\epsilon_\rho)^{1/4}}\approx-\frac{\epsilon_\rho}{4} \nonumber \\
\epsilon_{\rm CMB}&=-\frac{\epsilon_\rho}{1+\epsilon_\rho}\approx-\epsilon_\rho, \nonumber
\end{align}
where in the rightmost expressions we also give the small distortion approximations.
Here, $\Delta T_{\rm in} = T_{\rm in} - T_{\rm CMB}$. To compute $\epsilon_\rho\equiv\epsilon_\rho(\epsilon, \ma, \Tin)$ in the general case we can use 
\begin{align}
\epsilon_\rho&=-\frac{1}{G_3}\int \xin^3 \,P\left(\epsilon, \ma, \Tin \xin  \right)\,n_{\rm bb}(\xin)\id \xin,
\end{align}
where the conversion probability is evaluated at $\omega =\Tin \xin$. 
With Eq.~\eqref{eq:Pcon_simp} and the definition for $\gammacon$, we can rewrite the conversion probability as
\begin{align}
P\left(\epsilon, \ma, \Tin \xin\right)&=1-\exp\left(- \frac{\gamma_{\rm con}\xin \Tin}{\TCMB(\zcon)}\right) \nonumber \\ 
&\equiv P\left(\gammacon^*, \xin\right)
\end{align}
with $\gammacon^*=\frac{\gammacon\Tin}{\TCMB(\zcon)}$. It is therefore possible to determine the initial temperature by solving the equation
\begin{align}
\frac{\gammacon}{\gammacon^*}&=\left[1+\epsilon_\rho(\gammacon^*)\right]^{1/4}
\end{align}
for $\gammacon^*$. No analytic form for $\gamma^*_{\rm con}$ exists, so we obtain the solutions numerically.
Assuming small conversion probability (i.e., $\gamma^*_{\rm con}\ll 1$), we find\footnote{This approximation is obtained by assuming that all contributions to the integral come from $\xin<\gammacon^{-1}$ such that $P\left(\gammacon^*, \xin\right)\approx \gammacon\xin-\frac{1}{2}\gammacon^2\xin^2$.}
\begin{align}
\epsilon_\rho(\gammacon^*)&\approx
-\gammacon^*\,\frac{G_4}{G_3}\left[1-\frac{\gammacon^*}{2}\,\frac{G_5}{G_4}\right].
\end{align}
%
\begin{figure}
\includegraphics[width=\columnwidth]{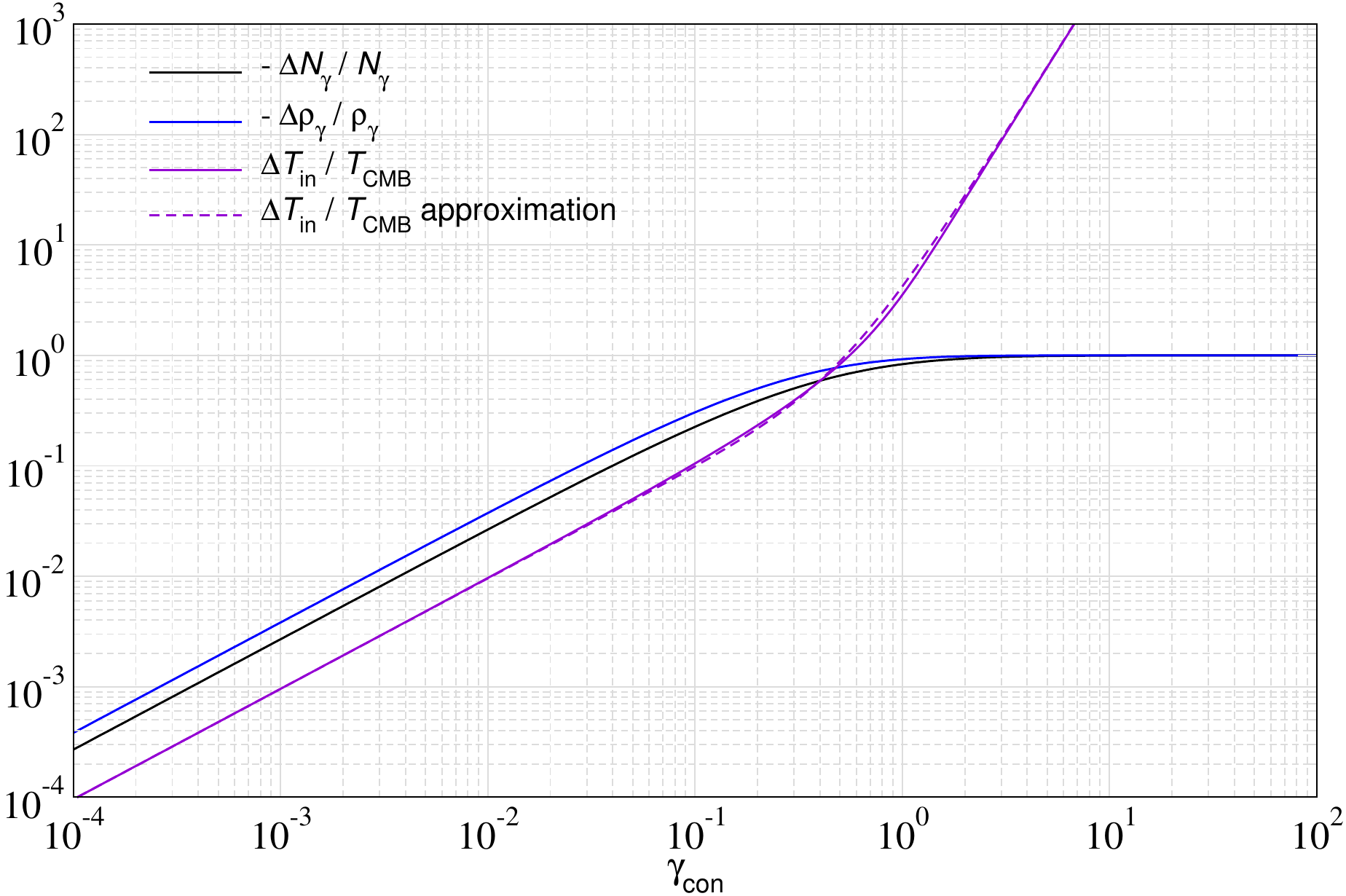}
\caption{Dependence of $\Delta \rho_\gamma/\rho_\gamma\big|_{\rm con}$, $\Delta N_\gamma/N_\gamma\big|_{\rm con}$ and $\Delta \Tin/\TCMB$ on $\gammacon$. We also show the approximation in Eq.~\eqref{eq:Tin_approx}}.
\label{fig:PC_functions_gamma}
\end{figure}
%
Keeping all terms up to second order in $\epsilon_\rho$ and $\frac{\Delta\Tin}{\TCMB}$ we can also show that the initial temperature is determined by
\begin{align}
\frac{\Delta\Tin}{\TCMB}
&\approx \frac{\gamma_{\rm con}}{4}\,\frac{G_4}{G_3}\left[1+
\frac{7}{8}\left(\frac{G_4}{G_3}-\frac{4}{7}\frac{G_5}{G_4}\right)\,\gamma_{\rm con}\right]
\nonumber
\\[1mm]
&\approx 0.9581\,\gamma_{\rm con}+0.8627\,\gamma^2_{\rm con}.
\end{align}
with good convergence at $\gamma_{\rm con}\lesssim 0.2$. For $\gamma_{\rm con}\lesssim 100$, we find 
\begin{align} \label{eq:Tin_approx}
\frac{\Delta\Tin}{\TCMB}
&\approx \frac{\gamma_{\rm con}}{4}\,\frac{G_4}{G_3}\,\bigg(1+3.3891 \gamma_{\rm con}^2\bigg)
\end{align}
to represent the full numerical result well, with largest departures of $\simeq 15\%$ around $\gammacon\simeq 1$. This can be useful for estimates, though in our main results we use the full numerical solution obtained from {\tt CosmoTherm}. 
%

In Fig.~\ref{fig:PC_functions_gamma} we illustrate the dependence of $\Delta \rho_\gamma/\rho_\gamma\big|_{\rm con}$, $\Delta N_\gamma/N_\gamma\big|_{\rm con}$ and $\Delta \Tin/\TCMB$ on $\gammacon$. For $\gammacon<1$, a linear scaling is found. For large $\gammacon$, both $\Delta \rho_\gamma/\rho_\gamma\big|_{\rm con}$ and $\Delta N_\gamma/N_\gamma\big|_{\rm con}$ approach $-1$, while $\Delta \Tin/\TCMB$ continues to rise as $\Delta \Tin/\TCMB \propto \gammacon^3$. In this limit, most of the initial photons are found deep in the Rayleigh-Jeans tail of the initial CMB blackbody with $\Tin>\TCMB$. We caution that corrections to the Landau-Zener formalism may become important in this regime \citep{Carenza2023}, and we reserve a more detailed treatment for future work.

\subsection{Distortion in the $\mu$ and $y$-eras ($10^3\lesssim z \lesssim \pot{2}{6}$)}
\noindent Following \citet{Chluba2024DP}, we can cleanly define the distortion shape that is introduced in the late phase of the evolution, when Comptonization is mostly negligible. We start with the initial spectrum at first order in small quantities as 
\begin{align}
\label{eq:n_dp}
n(\xin)&\equiv n_{\rm bb}(\xin)\exp\left(-\frac{\gammacon \xin \Tin}{\TCMB}\right)
\nonumber \\
&\approx n_{\rm bb}(x)+G(x)\,\frac{\Delta \Tin}{\TCMB}- x \,n_{\rm bb}(x)\,\gammacon
\end{align}
at $x\lesssim \gammacon^{-1}$. For convenience, we mention the standard spectra 
\begin{subequations}
\begin{align}
G(x)&=\frac{x \expf{x}}{(\expf{x}-1)^2},
\\
Y(x)&=G(x)\big[x\coth(x/2)-4\big],
\\
M(x)&=G(x)\left[\frac{1}{\beta_M}-\frac{1}{x}\right],
\end{align}
\end{subequations}
with $\beta_M=2.1923$. Heuristically, $G(x)$, $Y(x)$, and $M(x)$ describe the spectral shape of temperature shifts, $y$-distortions, and $\mu$-distortions, respectively. Using $\Delta \Tin/\TCMB\approx G_4 \gammacon/[4 G_3]$, we can write the distortion with respect to the CMB blackbody as 
\begin{align}
\label{eq:Dn_a}
\Delta n(x)&\equiv
n(\xin)-n_{\rm bb}(x)
\nonumber \\
&\approx \left[\frac{G_4}{4\,G_3}\,G(x)-x\,n_{\rm bb}(x)\right]\,\gammacon
\nonumber \\ 
&\equiv \left\{\left[\frac{G_4}{4\,G_3}-\frac{G_3}{3\,G_2}\right]\,G(x)
+A(x) \right\}\gammacon.
\end{align}
In the last line, we separated out the photon number carrying term to define the axion distortion spectrum
\begin{align}
\label{eq:Dn_a_N}
A(x)
&=\left[\frac{G_3}{3\,G_2}\,G(x)-x\,n_{\rm bb}(x)\right]
\end{align}
in a photon number-conserving way, i.e., $\int x^2 A(x)\id x=0$. In the small distortion limit, the typical spectral distortion that is {\it initially} formed is given by $A(x)$, almost independent\footnote{For $\gammacon \gtrsim 0.01$ some corrections from higher order frequency terms are noticeable but we confirmed that at $\epsilon \lesssim 10^{-7}$ these become subdominant.} of the value of $\gammacon \lesssim 0.01$. For axion models with the same value for $\gammacon$, differences in the final distortion shape and constraints can thus only arise from the variation of the axion mass, which directly controls the conversion redshift. 
%
\begin{figure}
\includegraphics[width=\columnwidth]{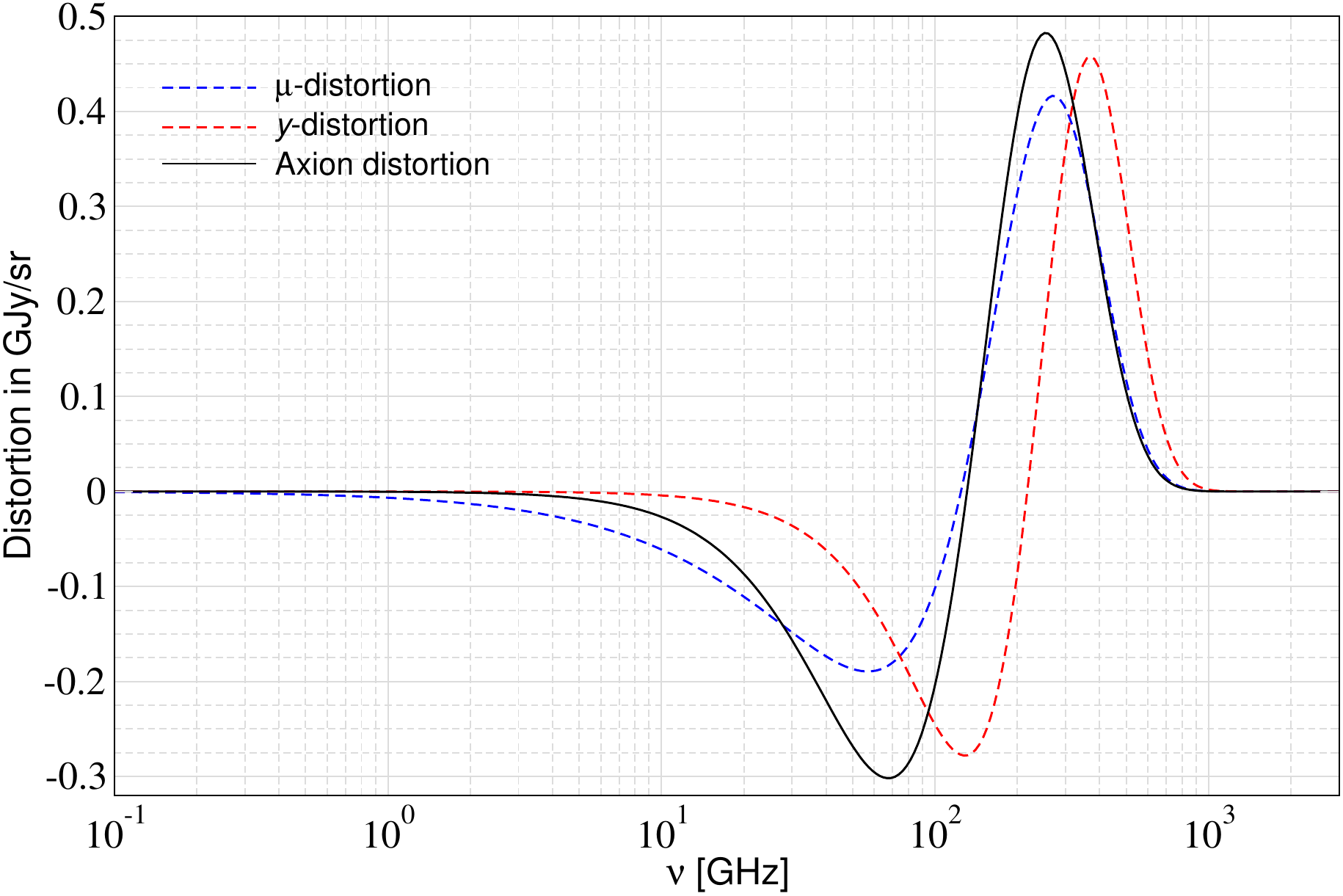}
\caption{Normalized distortion spectra $M^*(x)=1.401\,M(x)$, $Y^*(x)=Y(x)/4$ and $A^*(x)=-4.3354\,A(x)$ [with $4.3354=1/0.23066$]. The axion distortion has a new shape, with a null close to that of the $\mu$-distortion. Note that $A(x)$ has a negative sign in comparison to the $\mu$ and $y$ distortions.}
\label{fig:A_distortion}
\end{figure}
%
For the energy integral, we find $$\int x^3 A(x)\id x=\frac{4 G^2_3}{3 G_2}-G_4\approx -0.23066 \,G_3.$$
For comparison, one has the integrals $\int x^3 M(x)\id x=G_3/1.401$ and $\int x^3 Y(x)\id x=4 G_3$ for a $\mu$ and $y$-distortion respectively. In Fig.~\ref{fig:A_distortion} we show the (energy) normalized distortion spectra such that $\int x^3 A^*(x)\id x=G_3$, $\int x^3 M^*(x)\id x=G_3$ and $\int x^3 Y^*(x)\id x=G_3$. The axion distortion is initially neither close to the spectrum of a $\mu$ nor $y$-distortion. By equating the related energy densities we have
\begin{align}
\label{eq:mu_eff}
\gammacon\,\int x^3 A(x)\id x=\mu^*\int x^3 M(x)\id x
\end{align}
which implies $\mu^*= -0.3231\,\gammacon$. This effective $\mu$-parameter is in agreement with the estimates from Sect.~\ref{sec:level2}. It also indicates that the simple distortion energy density based constraints, Eq.~\eqref{eq:limit_total}, can be expected to work extremely well.

For the dark photon distortion, we found that the signal is extremely close to that of a $\mu$-distortion \citep{Chluba2024DP}. In this case, the distortion shape did not depend much on the conversion redshift. In contrast, we expect some evolution of the axion distortion due to Compton scattering in the $\mu$ and transition eras, since the distortion shape is not close to an equilibrium spectrum. We confirm this explicitly using {\tt CosmoTherm}.

\subsection{Late evolution ($z \lesssim 10^3$)}
\noindent The late evolution for the axion distortion is more complicated due to the fact that multiple conversions can occur (see Fig.~\ref{fig:x_gamma_crit}). However, Comptonization becomes inefficient at $z\lesssim 10^3$, simplifying the treatment. In the small distortion limit ($\gammacon \ll 1$), one then has the solution
\begin{align}
\label{eq:multiple_approx}
\Delta n(x)\approx 
\frac{\Delta T}{T}\,G(x)-n_{\rm bb}(x) \, P_{\rm \gamma_{||} \rightarrow  a}(\gammacon, x),
\end{align}
where $G(x)=x \expf{x}/(\expf{x}-1)^2$ and $$\frac{\Delta T}{T} \approx \frac{1}{3 G_2}\int x^2 n_{\rm bb}(x) \,P_{\rm \gamma_{||} \rightarrow  a}(\gammacon, x)\id x.$$
This term simply follows by demanding that the distortion spectrum carries no photon number, $\int x^2 \Delta n(x) \id x=0$, and ignores a small (unobservable) change of the average CMB temperature. We will show below that this approximation indeed captures all the main effects at late times ($z\lesssim 10^4$) and can thus be reliably used for deriving distortion constraints.

\section{CMB Spectral Distortion solutions}
\label{sec:level4}
\noindent We now illustrate the full evolution of the distortion for representative cases by utilizing the numerical solver \texttt{CosmoTherm}. We will start with the evolution in the large distortion regime to illustrate the differences with respect to the dark photon distortions computed in \citet{Chluba2024DP}. We then consider cases in the $y$ and $\mu$-era and briefly discuss the late evolution in the post-recombination era. In contrast to the dark photon case, the distortion evolves slightly, given that it is an out-of-equilibrium signal. Finally, we consider cases with multiple conversions demonstrating the validity of the approximations.

%
\begin{figure}
\includegraphics[width=\columnwidth]{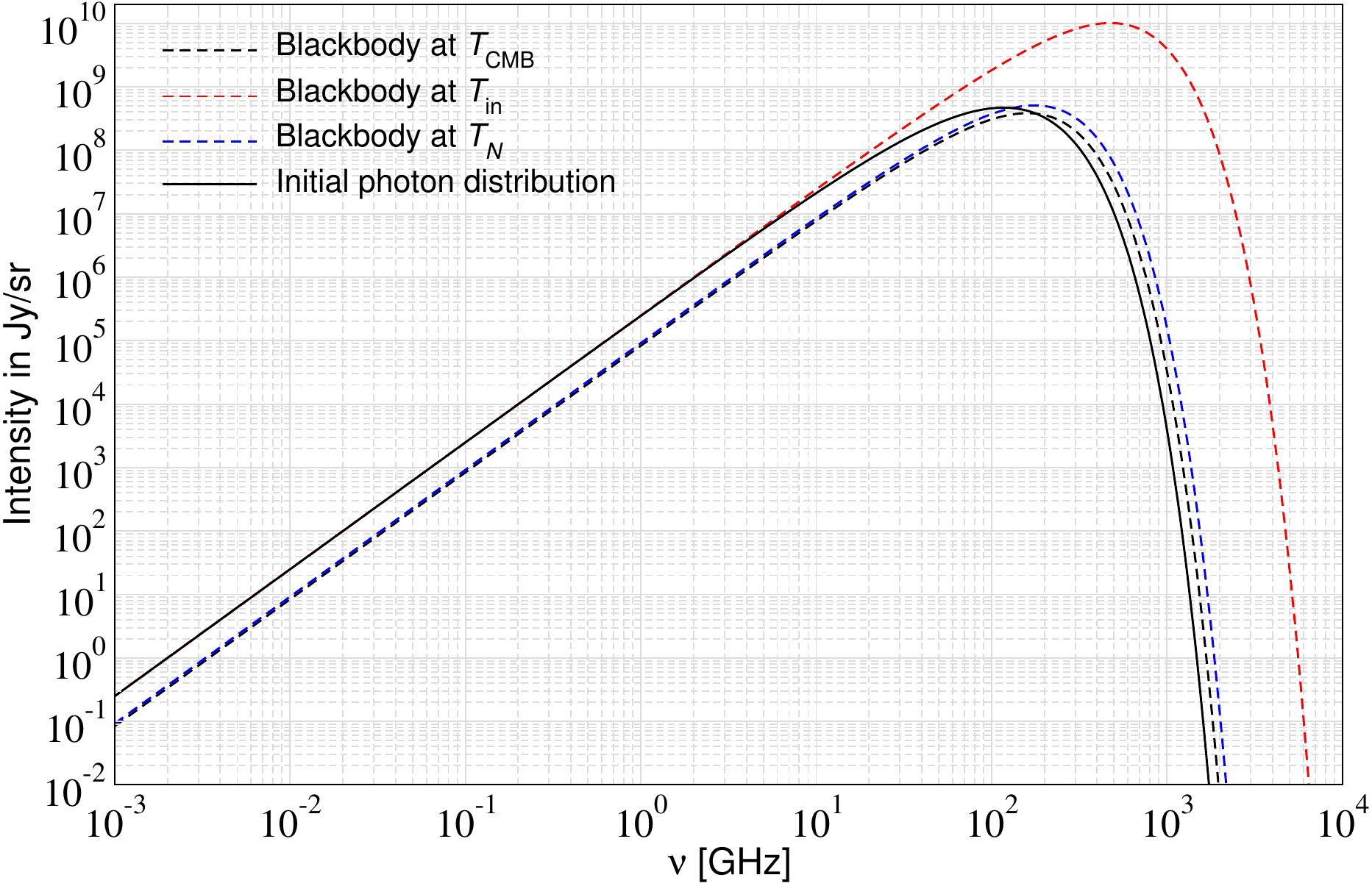}
\caption{Illustration of the initial spectrum for $\ma=10^{-4}\,\eV$ and $\epsilon=1.87$. See text for detailed discussion of this large distortion case.}
\label{fig:Initial_large}
\end{figure}
%
\subsection{Evolution in the large distortion regime}
\label{sec:large_evol}
\noindent We shall start by considering scenarios in the large distortion regime. One illustrative case is $\ma=10^{-4}\,\eV$ and $\epsilon=1.87$, implying $\zcon\approx \pot{3.21}{6}$ and $\gammacon\approx 0.784$. This model is already ruled out by \COBEF data as it leads to $\Delta \rho_\gamma/\rho_\gamma\big|_{\rm con}\approx -0.987$ and $\Delta N_\gamma/N_\gamma\big|_{\rm con}\approx -0.950$, which does not thermalize until today. The initial temperature has $\Delta \Tin/\TCMB\approx 1.97$, implying that the initial blackbody is $\approx 3$ times hotter than the CMB would be today. One reason for choosing this extreme situation is to elucidate the differences between the various distributions.

The initial spectrum is shown in Fig.~\ref{fig:Initial_large}. We can see that in comparison with the CMB blackbody, the initial photon distribution has a deficit in the Wien-Tail and an excess in the Rayleigh-Jeans part. The total photon distribution has an excess of photons over the CMB, which leads to very interesting non-linear dynamics: Compton scattering alone would drive the solution towards a Bose-Einstein distribution with a negative chemical potential at high frequencies. At low frequencies, stimulated Compton scattering leads to condensation of photons towards low frequencies, where these are efficiently destroyed by Bremsstrahlung and double Compton processes \citep{Levich1969, Khatri2011BE}. 
%
\begin{figure}
\includegraphics[width=\columnwidth]{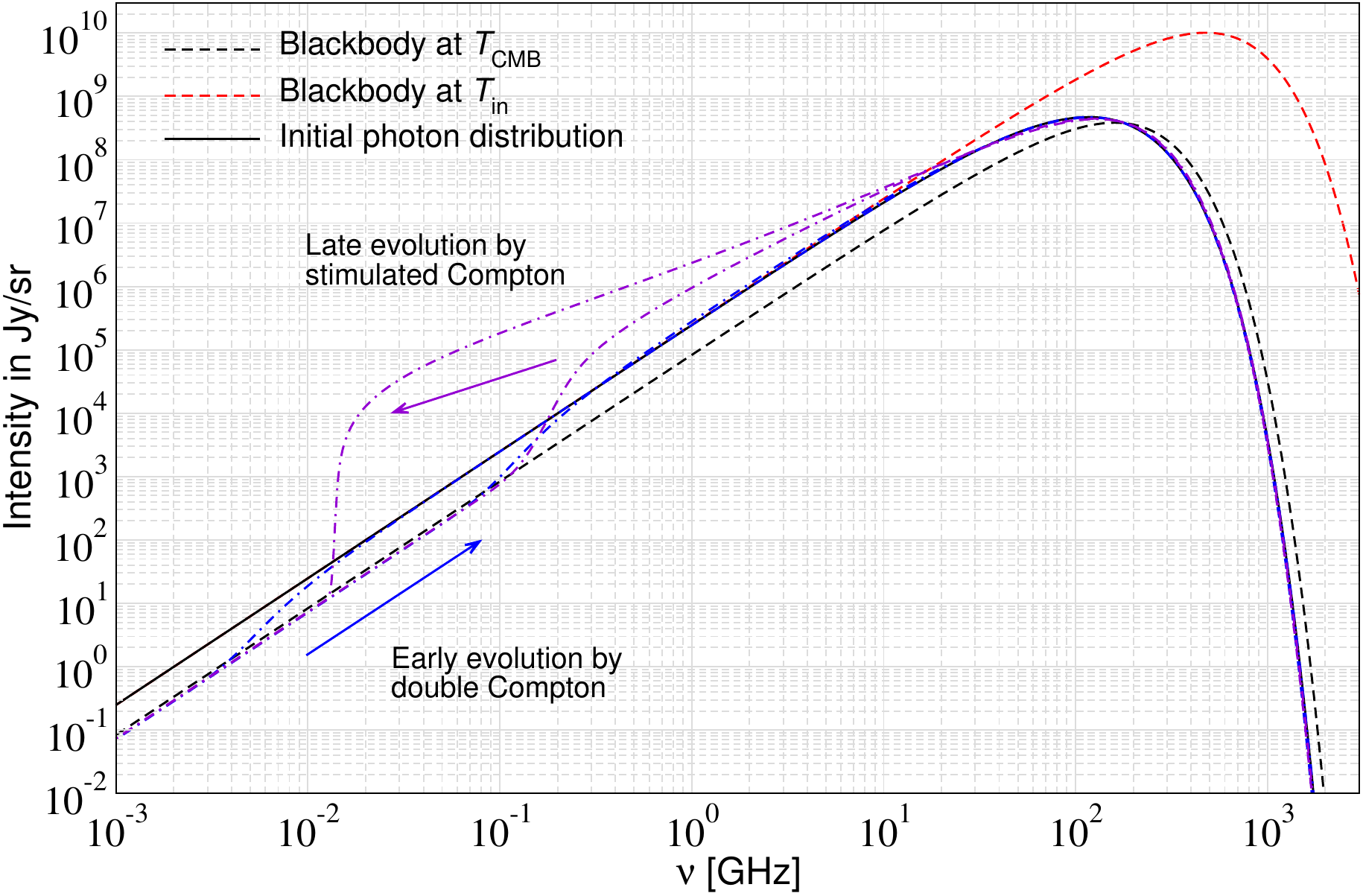}
\caption{Illustration for the evolution of the spectrum for $\ma=10^{-4}\,\eV$ and $\epsilon=1.87$. A photon shock develops at low frequencies for the considered parameters. Early intermediate states are in blue, later stages in purple.}
\label{fig:Initial_large_shock}
\end{figure}
%
Some of the evolutionary stages are illustrated in Fig.~\ref{fig:Initial_large_shock}. We only solved the problem up to a scattering $y$-parameter $\simeq 0.3$, since for the chosen parameters a strong photon shock \citep{Levich1969} develops at low frequencies, rendering our numerical treatment insufficient.

%
\begin{figure}
\includegraphics[width=\columnwidth]{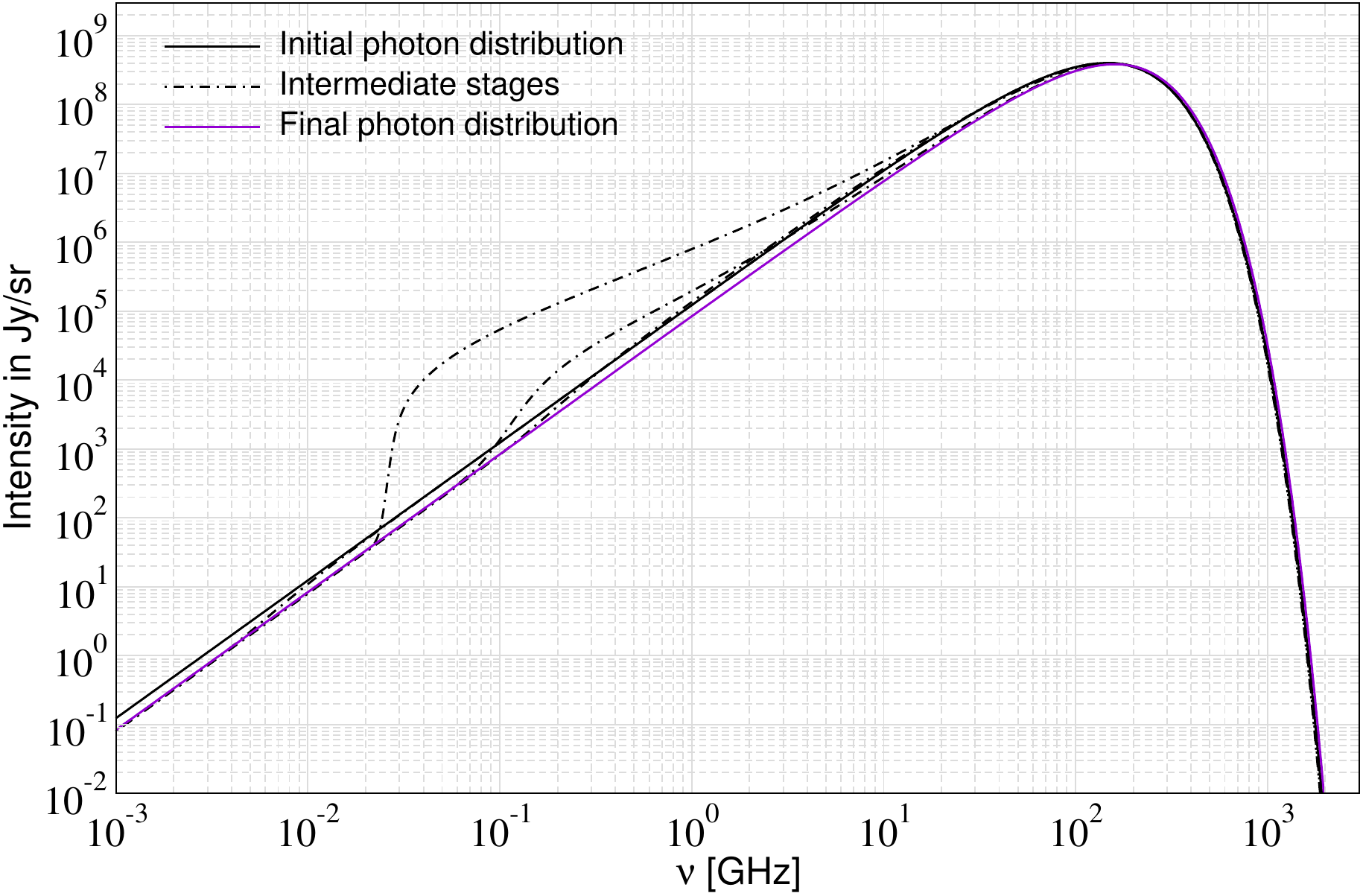}
\\[1mm]
\includegraphics[width=\columnwidth]{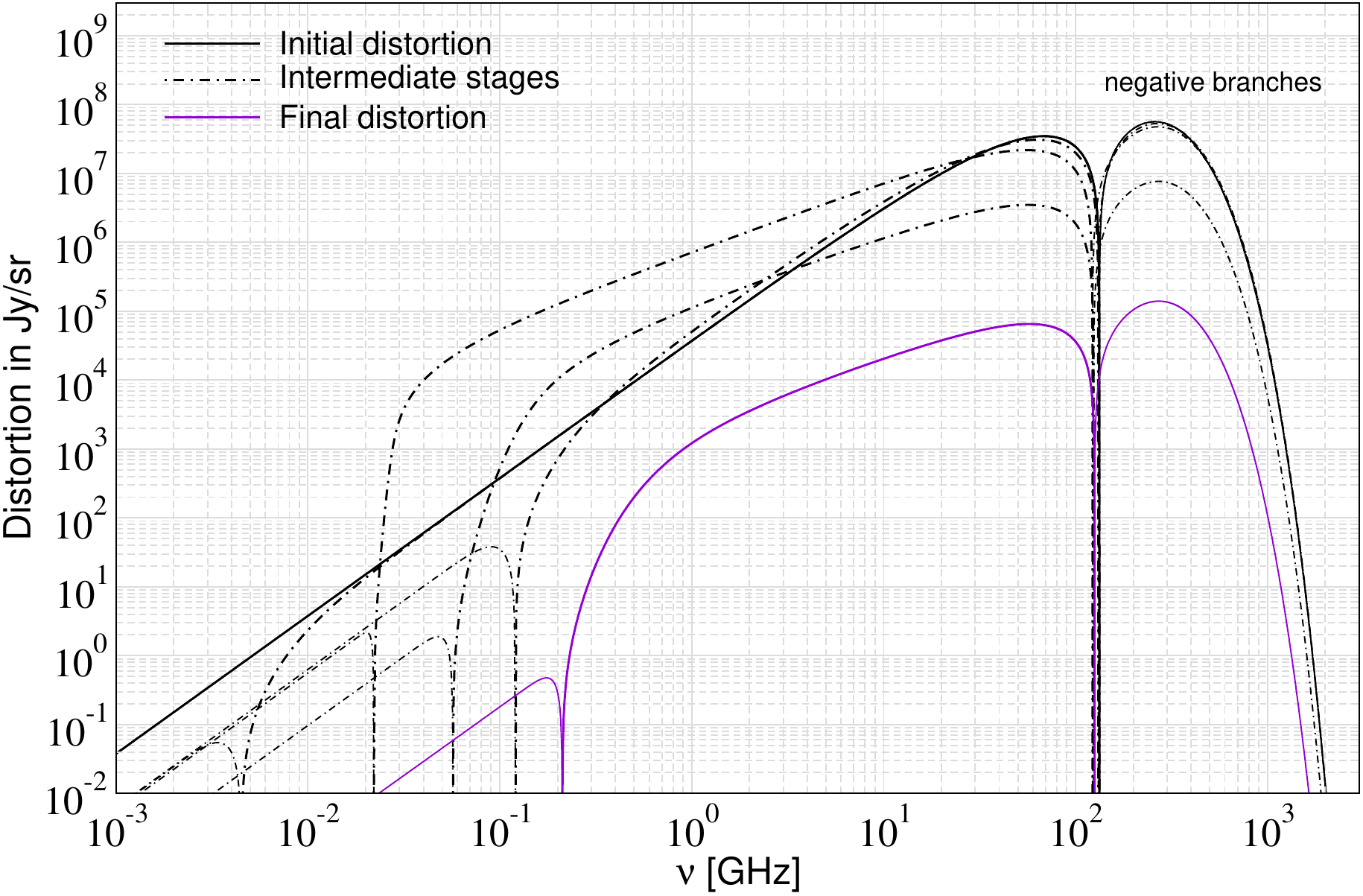}
\\
\caption{Illustration of distortion evolution for $\ma=10^{-4}\,\eV$ and $\epsilon=1.25$. The top panel shows the total photon spectrum, while the lower panel highlights the distortion. Negative parts of the distortion are shown as thin lines.}
\label{fig:Initial_large_stages}
\end{figure}
%
In Fig.~\ref{fig:Initial_large_stages}, we illustrate some of the evolutionarly stages for a less extreme scenario with $\ma=10^{-4}\,\eV$ and $\epsilon=1.25$, implying $\zcon\approx \pot{3.21}{6}$ and $\gammacon\approx 0.349$. Again, this model is already ruled out by \COBEF data as it leads to $\Delta \rho_\gamma/\rho_\gamma\big|_{\rm con}\approx -0.793$ and $\Delta N_\gamma/N_\gamma\big|_{\rm con}\approx -0.666$, which does not thermalize until today. The initial temperature is $\Delta \Tin/\TCMB\approx 0.483$, implying that the initial blackbody is $\approx 1.5$ times hotter than the CMB would be today. 
Even if this setup is very similar to the large distortion scenario discussed in Fig.~5 of \citet{Chluba2024DP}, the distortion evolution differs drastically. The most important feature is the drift of photons towards low frequencies by stimulated Compton scattering \citep{Chluba2008d}, as we already explained above.

\subsection{Evolution across the $\mu$ and $y$-eras ($10^3\lesssim z \lesssim \pot{2}{6}$)}
\noindent We now show the evolution of distortions from the $\mu$ to the $y$ era. For this we consider the small distortion regime. For illustration we shall use $\epsilon \approx \pot{4}{-2}$, which is a bit above the \COBEF limit for most of the relevant masses.

%
\begin{figure}
\includegraphics[width=\columnwidth]{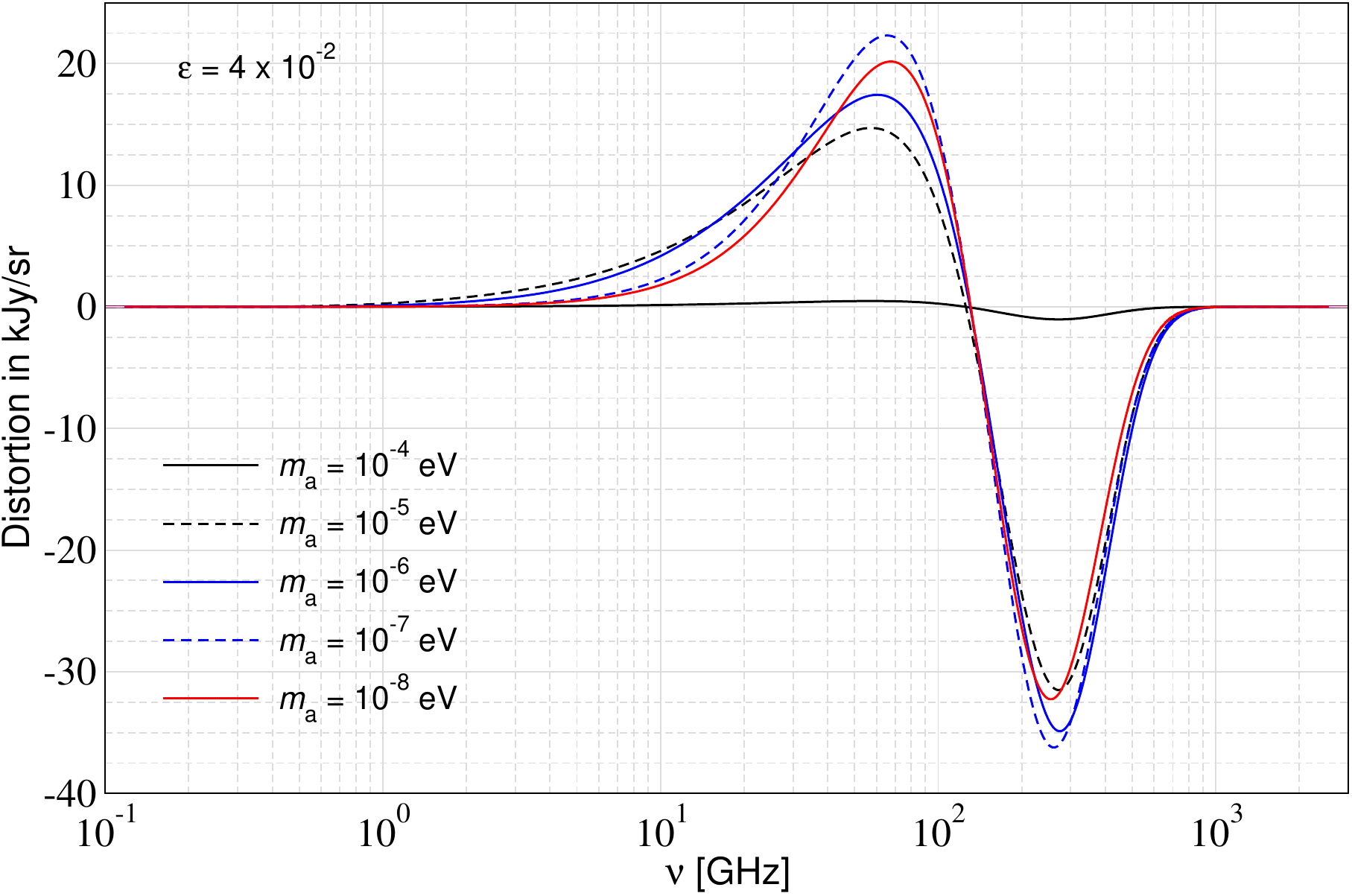}
\caption{Final distortion with respect to a blackbody for $\epsilon \approx \pot{4}{-2}$ and varying values of $\ma$. The final spectrum is evaluated at $z=0$. For an axion mass of $\ma=10^{-4}\,\eV$, the distortion is almost fully thermalized (i.e., has a tiny amplitude) and has a shape that is very close to a $\mu$-type distortion. For $\ma\lesssim 10^{-8}\,\eV$, the distortion shape is close to the axion distortion, $A(x)$, defined in Eq.~\eqref{eq:Dn_a_N}.
\vspace{-3mm}
}
\label{fig:Distortion_var_mdp}
\end{figure}
%
The results for the final distortion and varying axion mass are shown in Fig.~\ref{fig:Distortion_var_mdp}. Since the axion mass determines the conversion redshift, one can think of $\ma$ as a redshift label, with higher masses converting earlier. In contrast to the dark photon distortion discussed in \citet{Chluba2024DP}, the shape of axion distortion shows some noticeable dependence on the conversions redshift. In particular, the amplitude of the signal increases slightly in the CMB bands, which implies a slight tightening of the constraints is expected. For $\ma=10^{-4}\,\eV$ the distortion is almost fully thermalized (lower amplitude), with a shape close to that of a $\mu$-type distortion. The transition to the weakly Comptonized regime occurs around $\ma\lesssim 10^{-8}-\pot{3}{-6}\,\eV$, with the shape of the distortion essentially freezing to the axion distortion, $A(x)$, defined in Eq.~\eqref{eq:Dn_a_N} at $\ma\lesssim 10^{-8}\,\eV$.

\subsection{Late evolution and effects of multiple conversions}
\noindent From Fig.~\ref{fig:x_gamma_crit}, we already concluded that multiple conversions are only expected to affect the distortion shape for axions with masses $\ma\lesssim \pot{\rm few}{-10}\,\eV$. For these masses, conversions occur at $z\lesssim 10^{3}$ such that Compton scattering corrections are already rather minor \citep[e.g.,][]{Chluba2015GreensII}, and the analytic solution in Eq.~\eqref{eq:multiple_approx}, which neglects Compton scattering and also corrections due to free-free absorption at low frequencies, is expected to work extremely well. 

To numerically confirm this statement, we have to treat multiple conversions with {\tt CosmoTherm}. For this we use the following procedure. Once the mass of the axion is fixed, we compute the conversion redshifts as a function of frequency, $x=\omega / \TCMB$. This then also allows us to compute $\gammacon(x)$ by summing the various contributions in redshift to obtain $P_{\gamma_{||} \rightarrow {\rm a}}(x)$. We assume that throughout, the distortion remains small such that corrections to the CMB temperature do not have any direct effect on the distortion. This is a very reasonable assumption, as for late time conversions we know that constraints from \COBEF restrict us to the small distortion regime.

%
\begin{figure}
\includegraphics[width=\columnwidth]{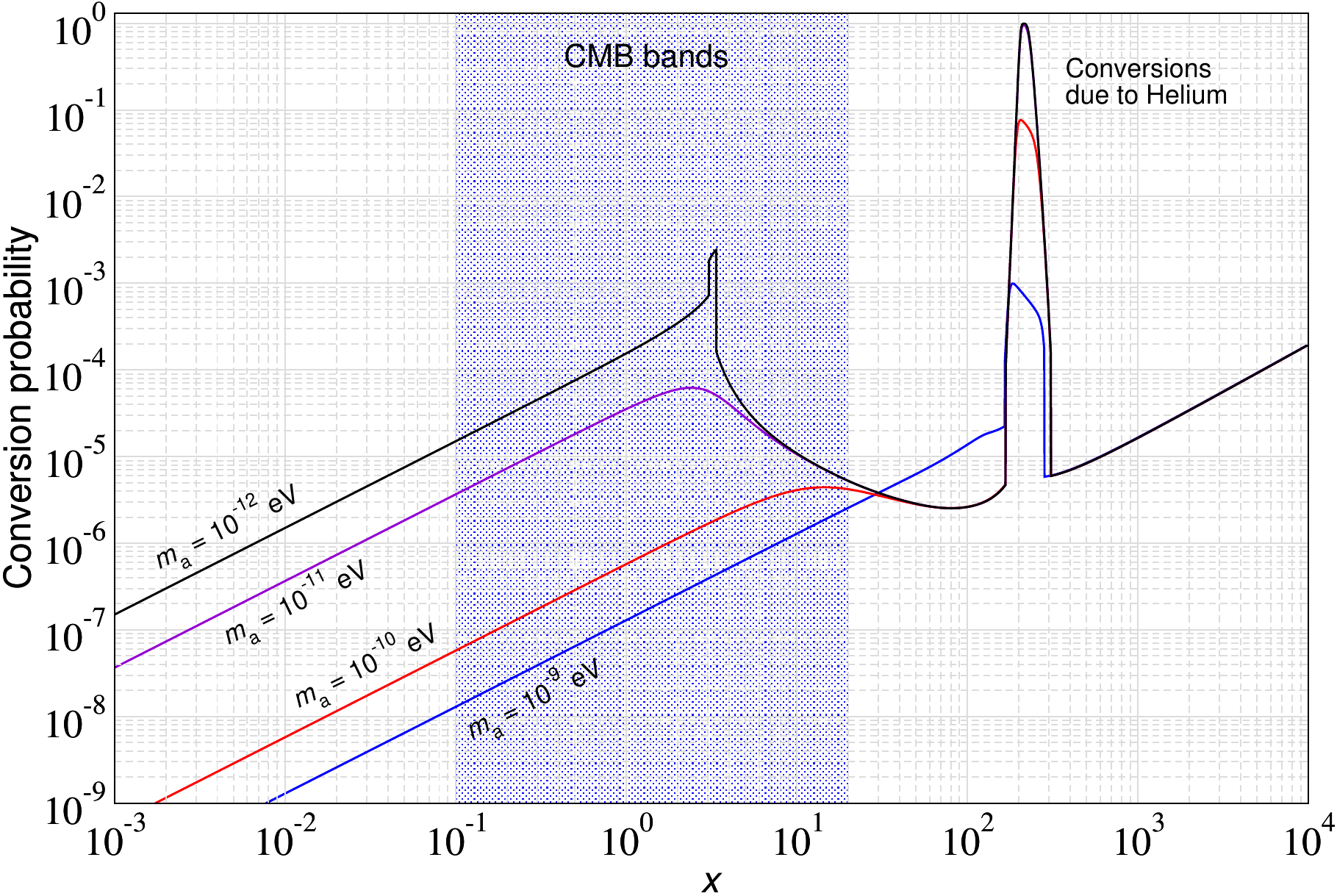}
\\[1mm]
\includegraphics[width=\columnwidth]{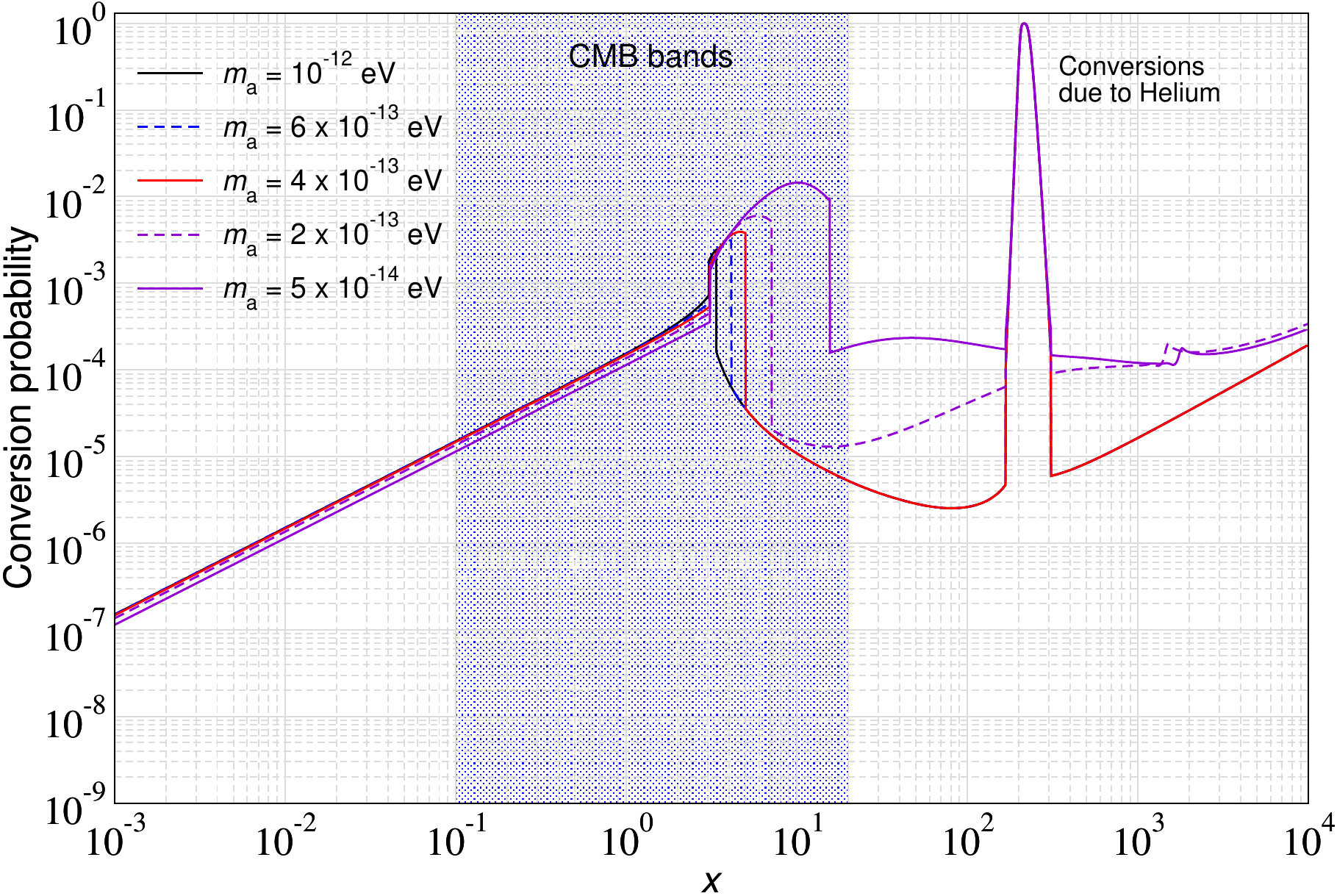}
\caption{Conversion probabilities for $\epsilon =10^{-3}$ and various axion masses. High-frequency structure becomes visible for $\ma\lesssim 10^{-10}\,\eV$. At axion masses $\ma\lesssim 10^{-12}\,\eV$, the conversion probability significantly increases again at high frequencies. Although a large spike is seen around $x \simeq 200$ due to Helium, very little of the CMB energy density is actually stored at these high frequencies, leaving no significant change in the the overall picture.}
\label{fig:P_x_var_m}
\end{figure}
%
To illustrate the effects, we show $P_{\gamma_{||} \rightarrow {\rm a}}(x)$ for a few examples in Fig.~\ref{fig:P_x_var_m}. Many features can be seen at high frequencies deep into the Wien tail of the CMB. At $\ma\lesssim 10^{-10}\,\eV$ we expect first that direct effects of conversions become noticeable for $x\lesssim 20$, although significant modifications are only expected for even smaller masses. We can also anticipate a change in the sign of the distortion for $10^{-12}\,\eV\lesssim \ma\lesssim 10^{-11}\,\eV$. This is because the conversion probability at high frequencies $x>3$ drops significantly for this mass range, implying that in the distortion the removal of photon number will take over and cause a net positive sign [as can be seen in Eq.~\eqref{eq:limit_total}] \citep{Chluba2015GreensII}. 

At $\ma\lesssim 10^{-12}\,\eV$, we observe a significant boost of the conversion probability at high frequencies. At $\ma<\pot{5}{-14}\,\eV$, we furthermore find no more conversions at low frequencies. We thus did not further consider these scenarios here, although we remark that conversions will still happen at $x>3$ for these low masses.

We note that previous works on axion-photon conversions in the context of CMB spectral distortions have treated the case of multiple conversions differently than us. For example, in the case of two conversions, both \citet{tashiro2013constraints} and \citet{Mukherjee2018} state that the conversion probability is $P(\gamma \rightarrow a) = p_{1}(1-p_{2}) + p_2(1-p_1)$ where $p_1$ and $p_2$ are essentially given by expression $p = {\rm exp}(-\gamma_{\rm con} x)$ evaluated at each resonant redshift. This expression conveys the fact that one should not double count photons which may have converted into axions, but naturally neglects any redistribution of photons in phase space between $z_{1}$ and $z_2$. Our numerical approach improves upon this by tracking the evolution of the photon occupation number at all times, including effects of energy redistribution through Compton interactions, as well as possible photon replenishment through efficient Bremsstrahlung processes at low frequencies.

\subsubsection{Late distortion solutions}
\noindent To confirm the validity of the analytic treatment, we implemented the multiple conversion scenario in {\tt CosmoTherm}. For this we assume that the total change in the CMB temperature is negligible and that we can simply set $\Tin\approx \TCMB$. The effect of this is that all timescales for photon interactions are slightly modified at a level of the relevant $\Delta \Tin/\TCMB$ (i.e., higher order). To present the distortion solutions, we remove any temperature shift terms by demanding the distortion fulfills $\int x^2 \Delta n_{\rm d}(x)\id x=0$. This is equivalent to saying that we compare the final spectrum to that of a blackbody at the number density based temperature, $T_N$ \citep{Chluba2015GreensII, Chluba2024DP}.

%
\begin{figure}
\includegraphics[width=\columnwidth]{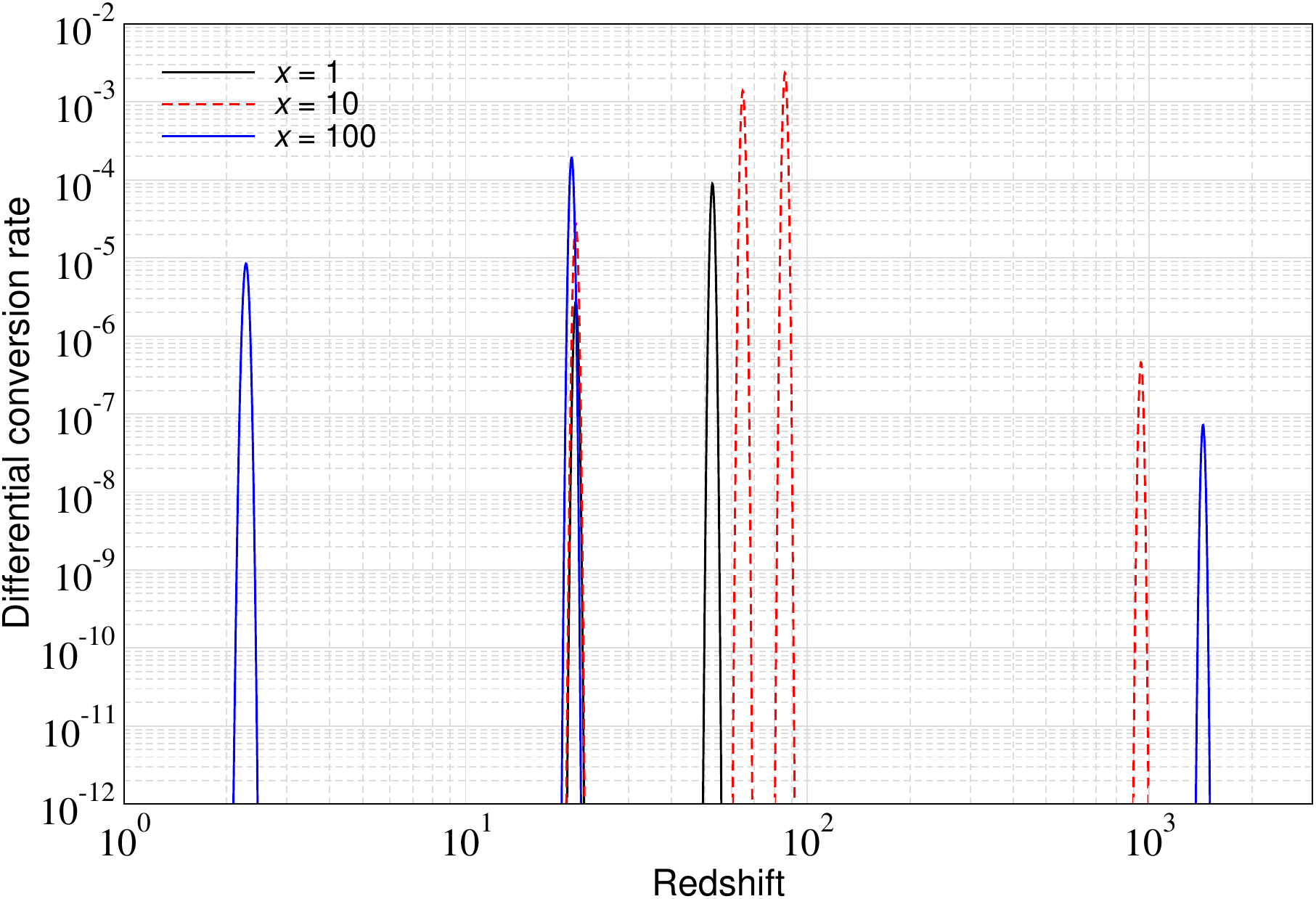}
\caption{Differential conversion probability for $\ma = 10^{-13}\,\eV$ and $\epsilon=10^{-3}$ and varying frequency. The conversion redshifts can be easily understood with Fig.~\ref{fig:x_gamma_crit}, giving three conversions for $x=1$, five for $x=10$ and three for $x=100$. Note that all frequencies convert at $ z\simeq 2$ for these parameters, with equal differential conversion rates. As a result, three overlapping contours appear for this low redshift conversion}
\label{fig:dP_dz_axion-new}
\end{figure}
%
To follow the multiple conversions in redshift and frequency, we add the photon conversion term 
\begin{align}
\frac{\id n(x, z)}{\id z}\Bigg|_{\rm con} = - \sum_{i}\gammacon(z_{{\rm con},i}, x) \,x \, \delta (z-z_{{\rm con},i})\,n(x, z) \nonumber
\end{align}
to the radiative transfer equation. In practice we replace the $\delta$-distribution by a narrow Gaussian, $G(z, z_{{\rm con},i})$, in redshift ($\Delta z/z \simeq 10^{-2}$). Once we have determined all the conversion redshifts we find the maximal and minimal conversions redshifts to ensure that inside this range the {\tt CosmoTherm} solver takes very fine redshifts steps to not miss any of the conversion loci. In Fig.~\ref{fig:dP_dz_axion-new} we illustrate the setup for $\ma = 10^{-13}\,\eV$ and $\epsilon=10^{-3}$ and varying frequency. We show the coefficient $\id P(x, z)/\id z=\sum_{i}\gammacon(z_{{\rm con},i}, x) \,x\,G(z, z_{{\rm con},i})$. The conversion redshifts can be easily understood with Fig.~\ref{fig:x_gamma_crit}, giving three conversions for $x=1$, five for $x=10$ and three for $x=100$. The conversions during neutral helium recombination are subdominant. We note that the relative importance of the conversions depends on the assumption that the comoving magnetic field strength is constant.

%
\begin{figure}
\includegraphics[width=\columnwidth]{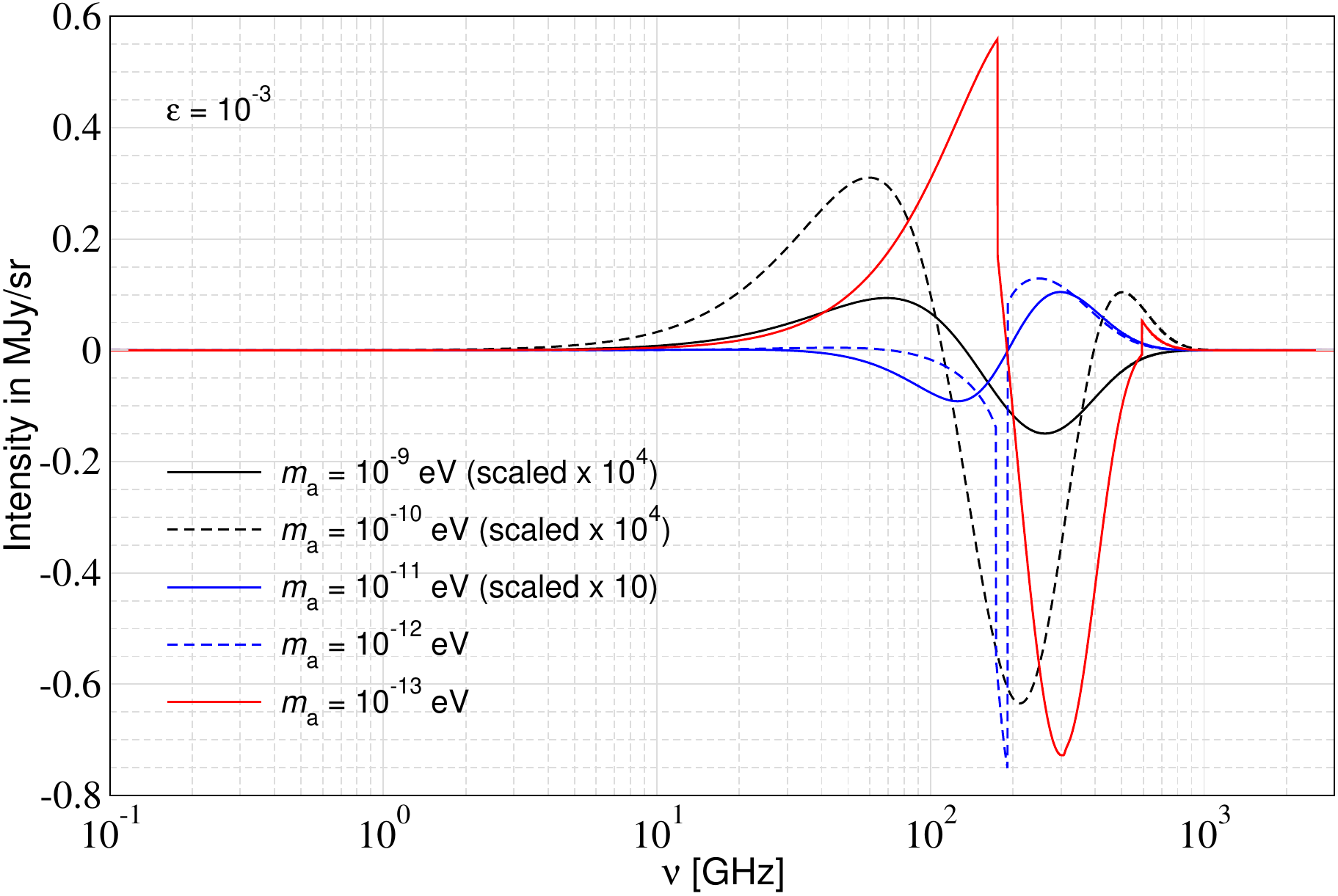}
\caption{Final distortion for $\epsilon=10^{-3}$ and varying masses. The complicated conversion structure leads complex spectral responses. Note that we scaled the distortions for $\ma=10^{-9}\,\eV$, $\ma=10^{-10}\,\eV$ and $\ma=10^{-11}\,\eV$ to make them more comparable.}
\label{fig:DI_var_m_axion}
\end{figure}
%
%
\begin{figure*}
\includegraphics[width=1.8\columnwidth]{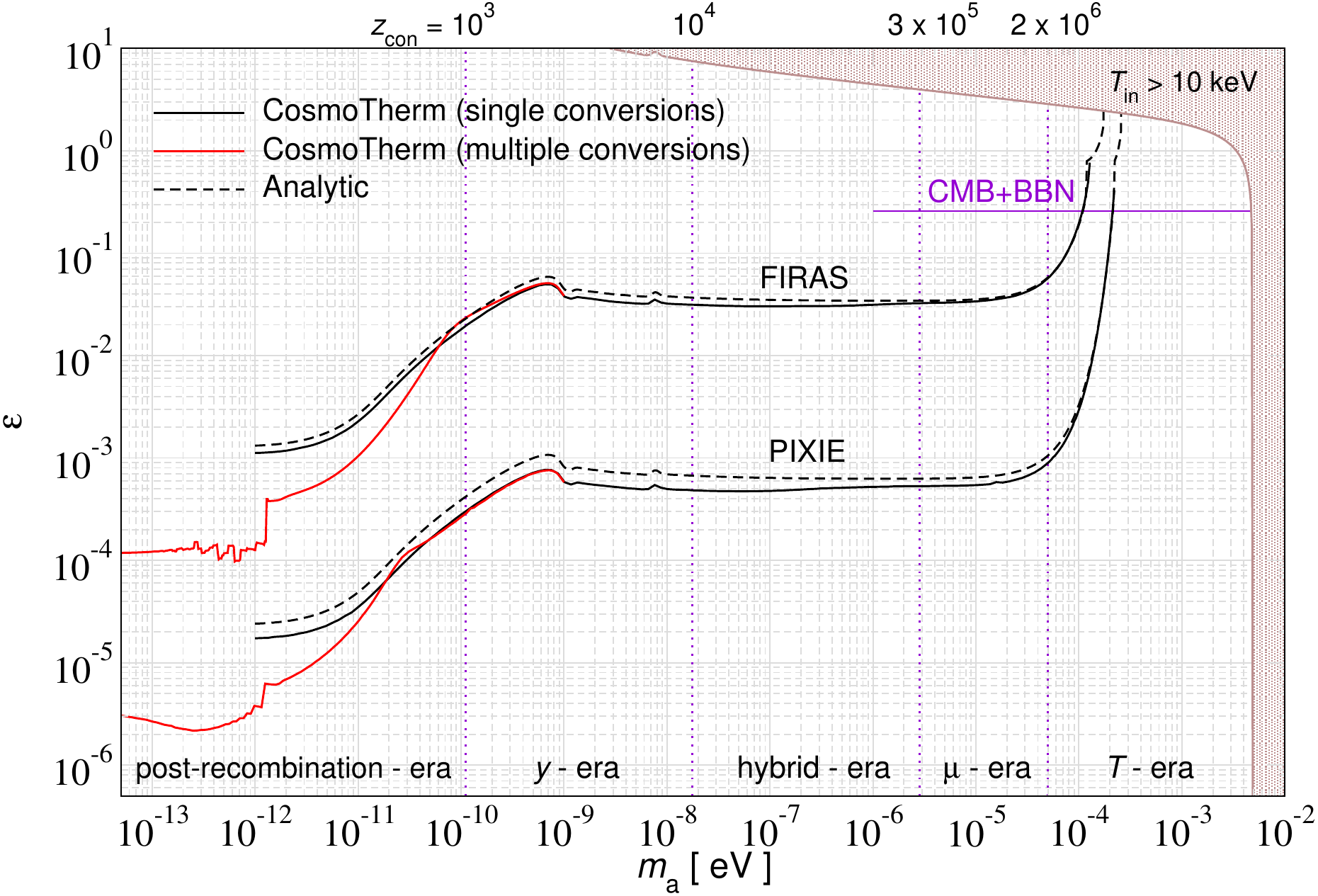}
 \caption{CMB spectral distortion limits (95\% c.l.) from \COBEF and \PIXIE. The solid lines show those limits obtained directly with {\tt CosmoTherm} while the dashed lines show the estimates based on Eq.~\eqref{eq:limit_total} in the small distortion limit. The red solid lines are obtained with {\tt CosmoTherm} using the multiple conversion setup. In the shaded region, the initial temperature at the conversion redshift exceeds $10\,\keV$. The purple line gives the estimated constraint from considerations of $N_{\rm eff}$.}
\label{fig:Limits_distortion}
\end{figure*}
%
In Fig.~\ref{fig:DI_var_m_axion} we illustrate the numerical solution for a few masses. We fixed $\epsilon=10^{-3}$, and scaled the distortions for $\ma=10^{-9}\,\eV$, $\ma=10^{-10}\,\eV$ and $\ma=10^{-11}\,\eV$ to make them more comparable. We find that all of our expectations are confirmed. In particular, the sign of the distortion is flipped for $\ma=10^{-11}\,\eV$ and $\ma=10^{-12}\,\eV$, as anticipated. 
We also compared the {\tt CosmoTherm} solution to the analytic approximation, Eq.~\eqref{eq:multiple_approx}, finding excellent agreement. 

We also note that we expect some corrections to the solutions from electron scattering effects. {\tt CosmoTherm} solves the Kompaneets equation, which in the presence of sharp spectral features does not capture all the physics of the photon redistribution process \citep[e.g.,][]{Chluba2012HeRec, Acharya2021}. However, at low redshifts, when the distortion does exhibit narrow features, the electron scattering probability is already extremely low such that we do not expect this to modify the distortion shapes significantly. A more careful study may consider this aspect in the future.

\section{CMB Spectral Distortion Limits}
\label{sec:level5}
\noindent We can now compute the distortion constraints from our treatment of the photon to axion conversion process. For details about the likelihood code we refer the interested reader to \citep{Chluba2024DP}. We evaluate the distortion limits for \COBEF \citep{Fixsen1996} and for a \PIXIE-like experiment \citep{Kogut2016SPIE} that can achieve a sensitivity to $|\Delta \rho/\rho | \lesssim \pot{2}{-8}$. A number of concepts that are in principle able to reach this sensitivity have been discussed by the community \citep[e.g.,][]{PRISM2013WPII, Chluba2021Voyage, SPECTER2024}.

In Fig.~\ref{fig:Limits_distortion} we present the distortion limits and forecasts on $\epsilon$ in Eq.~\eqref{eq:def_eps}. The solid lines are all directly obtained using {\tt CosmoTherm}, while the dashed lines are analytic approximations based on Eq.~\eqref{eq:limit_total}. For the {\tt CosmoTherm} runs we considered the single conversion treatments down to $\ma=10^{-12}\,\eV$. We also contrast these results to the multiple conversion treatment at $\ma\leq 10^{-9}\,\eV$. Modifications due to multiple conversion start becoming noticeable at $\ma\lesssim \pot{\rm few}{-10}\,\eV$, while otherwise the two treatments agree extremely well. We also confirmed that the analytic multiple conversion treatment gives consistent results at $\ma\lesssim 10^{-9}\,\eV$.  

For comparison, we also show the constraint based on comparing the CMB measurements of $N^{\rm CMB}_{\rm eff}=2.99\pm0.34$ (95\% c.l.) from \Planck+BAO \citep{Planck2018params} with the standard BBN value, $N^{\rm BBN}_{\rm eff}=3.046$ in Fig.~\ref{fig:Limits_distortion}. This yields $|\epsilon_\rho|<0.056$ (95\% c.l.) \citep{Chluba2024DP}, which implies $\epsilon <0.26$ (95\% c.l.) as shown. Given this bound, we did not push the {\tt CosmoTherm} computations very far into the high coupling regime, given the numerical challenges that stimulated scattering effects cause (see discussion in Sect.~\ref{sec:large_evol}). By comparing with the analytic approximation at the high mass end, we find the {\tt CosmoTherm} distortion constraints to not show any significant large distortion effects below the $N_{\rm eff}$ bound. This is an interesting feature of the axion conversion scenario, which we discuss below (see Sect.~\ref{sec:large_limits}). 

By comparing the analytic estimate in the single conversion case [i.e., based on Eq.~\eqref{eq:limit_total}] we can also see that the constraint is slightly underestimated for both \COBEF and \PIXIE. We attribute this to the slightly differing distortion shape of the axion distortion, $A(x)$, which has an enhanced amplitude at fixed energy density (see Fig.~\ref{fig:A_distortion}). For \PIXIE we find the full result from {\tt CosmoTherm} to be about $\simeq 30\%$ tighter, which we can anticipate from the ratio of the extrema of the $\mu$ and $A$ distortion spectra.

Finally, at the low mass end the constraints exhibit multiple edges and features. This is due to the complicated frequency-dependence of the distortion signal (see Fig.~\ref{fig:DI_var_m_axion}) and interplay with the distortion data. In our analysis, we did not include any band averaging, which will smooth some of these features. However, the small jump in the constraints at $\ma \simeq 10^{-12}\,\eV$ will likely remain and is due to the appearance of an increased conversion probability a $x>3$ with decreasing mass (see Fig.~\ref{fig:P_x_var_m}).

\begin{figure}
\includegraphics[width=\columnwidth]{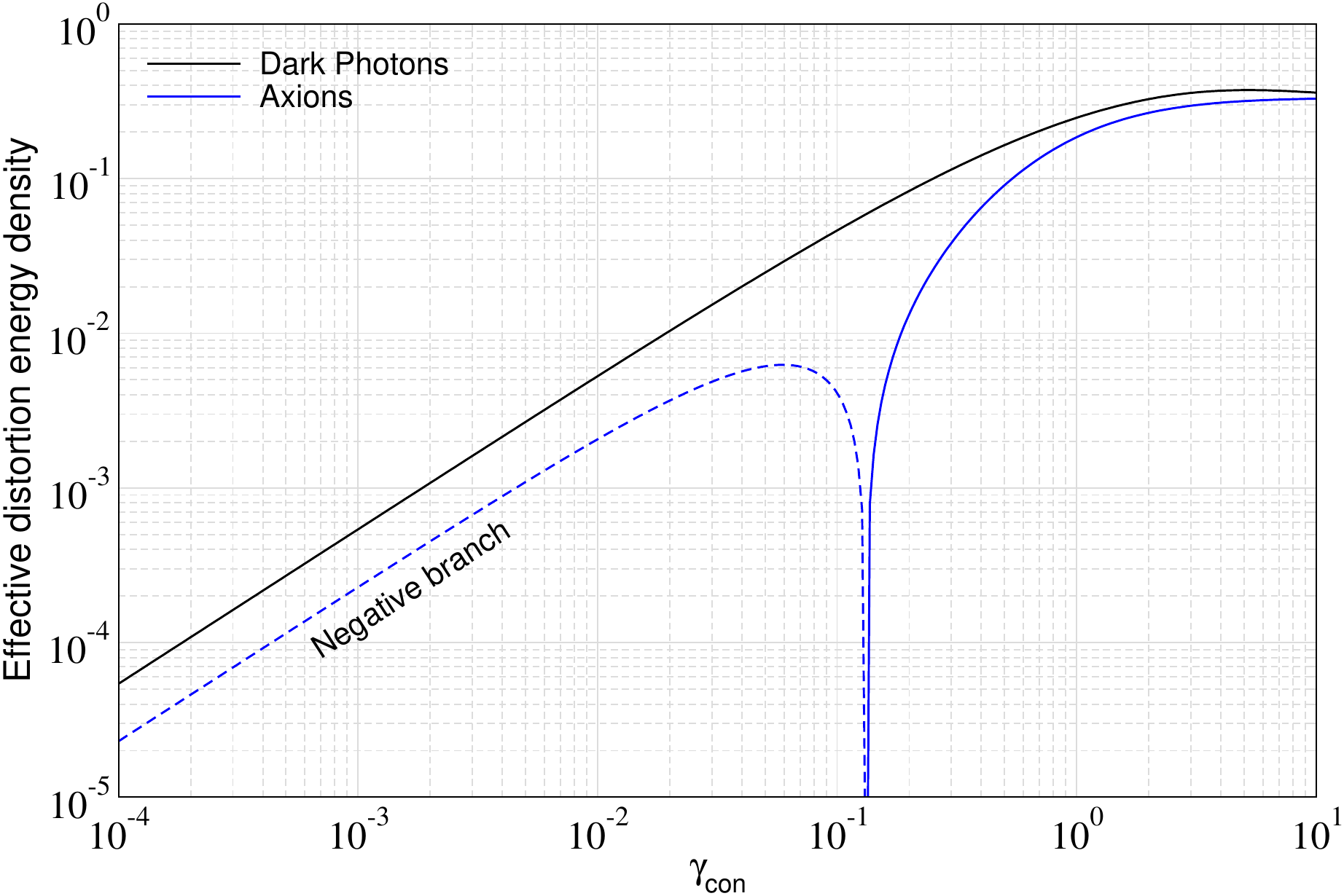}
\caption{Dependence of effective distortion energy density on $\gammacon$ for the axion and dark photon conversion cases. A sign flip is expected for the axion conversion distortion at $\gammacon \approx 0.13$ which is due to limitations of the approximation (see text for discussion).}
\label{fig:Drho_rho_distortion}
\end{figure}

\subsection{Constraints in the large distortion regime}
\label{sec:large_limits}
\noindent In our previous study of the dark photon conversion process we found that the analytic approximation based on the total distortion energy similar to Eq.~\eqref{eq:limit_total} underestimated the distortion constraints in the large distortion regime \citep{Chluba2024DP}. Here in contrast we find the analytic estimates to agree well up to the $N_{\rm eff}$ bound and then even predict some domain of modified limits (see slight kink at $\epsilon \simeq 0.8$ in Fig.~\ref{fig:Limits_distortion}), when the entropy extraction term in Eq.~\eqref{eq:limit_total} cancels the energy term\footnote{This was also seen in \citet{mirizzi2009constraining}.}. 

To understand this aspect better, in Fig.~\ref{fig:Drho_rho_distortion} we illustrate the dependence of the effective distortion energy density on $\gammacon$. The results were obtained numerically and illustrate the differences between the axion and dark photon conversion scenarios. For dark photons converting into photons, we find a net positive distortion energy density \citep{Chluba2024DP} for all $\gammacon$. In contrast, for the axions we have a net {\it negative} distortion energy density in the small distortion regime, which changes sign around $\gammacon \simeq 0.13$. This means that in the large distortion regime, we expect the distortion to flip sign in the axion scenario, which leads to the feature in the analytic limit (see Fig.~\ref{fig:Limits_distortion}).

\begin{figure}
\includegraphics[width=\columnwidth]{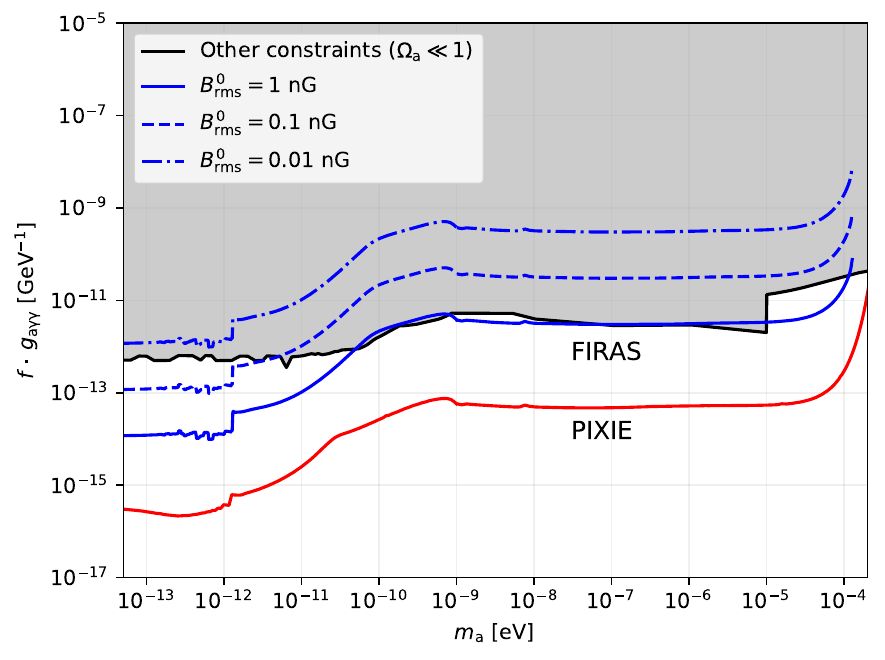}
\caption{Upper limits on $g_{\rm a \gamma \gamma}$ from CMB spectral distortions for different values of the large scale $B$-field. The factor $f$ represents a fudge factor discussed in the main text. The ``Other constraints" contour does not necessarily depend on $f$. A breakdown of these constraints can be found on the \href{https://cajohare.github.io/AxionLimits/docs/ap.html}{Github} of Ciaran O'Hare and also in \citet{OHare2024}.}
\label{fig:gagg_constraints}
\end{figure}

In our numerical computation with {\tt CosmoTherm} we do not find this behavior for a number of reasons. First, the approximation Eq.~\eqref{eq:limit_total} assumes small distortions, while the expected change of sign occurs in the large distortion regime, where the approximation becomes invalid. One could attempt to improve the approximation by considering more general expressions but this fails to capture a more significant aspect: for the dark photon conversion scenario the distortion always has a lack of photons. When entering the large distortion regime this bottlenecks the thermalization procedure because double Compton emission cannot easily replenish photons, and the distortion shape becomes altered in a way which inhibits thermalization, thus tightening the constraints over the small distortion approximations \citep{Chluba2020large}. In contrast, the axion distortion has an excess of photons even in the very large distortion regime\footnote{This can be demonstrated by computing the effective number density of photons based on the total energy density alone and comparing this with the actual photon number density, i.e., $(1+\epsilon_\rho)^{3/4}<(1+\epsilon_N)$.}. This leads to the condensation of photons towards low frequencies (e.g., see Fig.~\ref{fig:Initial_large_shock}) where they can be efficiently destroyed by double Compton absorption, entirely avoiding a large distortion bottleneck. Essentially, while in the large distortion regime, the photon number is efficiently adjusted until the small distortion regime is reached, at which point further evolution of the spectrum proceeds in the usual perturbative setup, yielding weak constraints. We leave a more detailed exploration of these interesting physical effects to the future.

\subsection{Limits on $g_{\rm a\gamma\gamma}$}
\label{sec:g_agg_limits}
\noindent When studying axion models, constraints are often placed in the parameter space of $g_{\rm a \gamma\gamma}$ versus $m_{\rm a}$. To do that here, we need to assume an amplitude for the transverse magnetic field. As discussed above, measurements coming from the \Planck satellite place constraints on the total root-mean-squared $B$-field value (as measured today). For scale-invariant fields this yields $B_{\rm rms}^0 = \sqrt{ \langle B^2 \rangle} \lesssim 1 \, {\rm nG}$ \citep{Ade:2015cva, Paoletti2022}. Future measurements by LITEBIRD of $B-$mode polarization \citep{LiteBIRD2024} aim to improve this bound by a factor of roughly $\simeq 1.5$. In contrast to the upper bounds set by CMB observations, recent results from \textit{Fermi}-LAT and H.E.S.S have refined our understanding of intergalactic magnetic fields, finding a corresponding lower limit of $B > 7.1 \times 10^{-7}$ nG with a coherence length of $\simeq 1$ Mpc \citep{HESS2023}. 

If magnetic fields exist on large scales, it is possible that they could been generated by seed fluctuations frozen in during inflation \citep{Durrer2013}. If this were the case, it is reasonable to expect that these magnetic fields could be present in the pre-/post-recombination plasma, with an amplitude that scales with redshift as $B(z) = B_0(1+z)^2$. Indeed, this was an assumption that went into the form of $\gamma_{\rm con}$ in Eq.~\eqref{eq:gamma_con_etc} as well as the constraints seen in Fig.~\ref{fig:Limits_distortion}. 
We note that this assumption does not seriously affect the interpretation of the constraints for masses $\ma\gtrsim \pot{\rm few}{-10}\,\eV$, as those distortion signals arise in single conversions. In this regime, one can thus find a corresponding mapping to the physical field strength at the respective conversion redshift. However, at lower masses, multiple conversions can occur and prevent this simplified remapping. In this situation, one can directly use the analytic solution for the spectral distortion, Eq.~\eqref{eq:multiple_approx}, to compute the distortion for a given magnetic field strength model, $B(z)$. Even for scale-invariant primordial magnetic fields, one expects a noticeable drop of comoving magnetic fields when crossing the recombination era \citep{Trivedi2018}.

Another aspect that enters the problem and interpretation of the constraints is the geometry and coherence length of the magnetic field. The precise geometry of a given magnetic field is quite model dependent, and is required to compute the transverse component $B^0_{\rm rms,T}$. Here, we package this in terms of a fudge factor $f$ which relates $B_{\rm rms,T} = f\, B_{\rm rms}^0$. For an isotropic bath of photons, simple angle-averaging arguments would suggest that $f = 1/3$ \citep{mirizzi2005photon, Mukherjee2018}\footnote{A factor of $2/3$ comes from averaging over solid angle, multiplied by an additional $1/2$ for the fact that only half the photon polarization states will be along $B_{\rm rms,T}$. }, though things become more complicated in the pre-recombination plasma due to rapid electron scattering interactions which can redistribute photon energies (Compton scattering events with energy exchange) and scramble polarization states (Thomson scattering events which lead to isotropization of the medium). As a result, we retain the $f$ factor in Fig.~\ref{fig:gagg_constraints} where we plot constraints in the $g_{\rm a \gamma \gamma}$ versus $m_{\rm a}$ parameter space.

\begin{figure}
\includegraphics[width=\columnwidth]{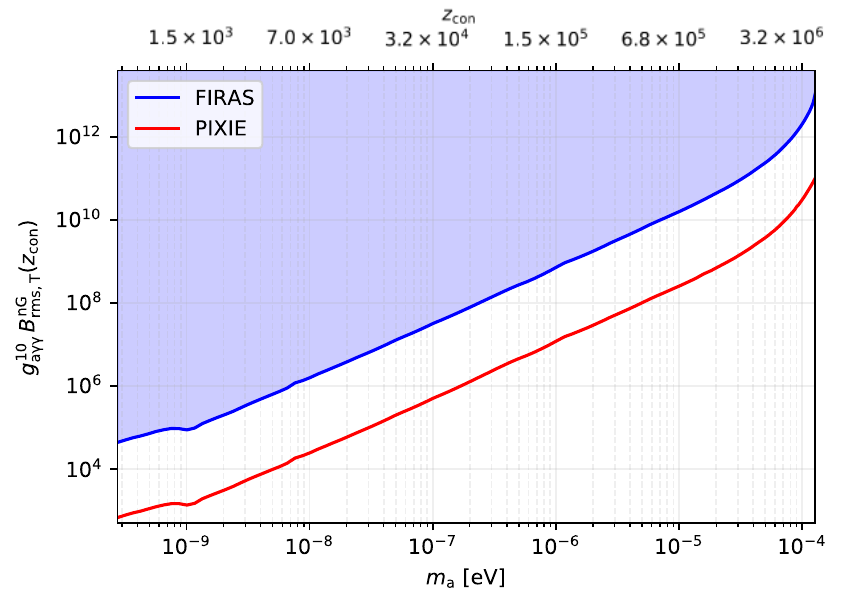}
\caption{Constraints on the model parameters valid for general (i.e. non-comoving) $B$-fields whose amplitudes should be evaluated at $B(z_{\rm con})$. Regions above the blue and red contours are constrained by present and future spectral distortion measurements. Here we restrict to mass ranges in which only a single resonant conversion is expected (e.g. with $z_{\rm con} > z_{\rm rec}$). Here we express $g_{\rm a \gamma \gamma}^{10} = g_{\rm a \gamma \gamma}/(10^{-10} \, {\rm GeV}^{-1})$ and $B_{\rm rms,T}^{\rm nG} = B_{\rm rms,T}/$nG. }
\label{fig:generic_B}
\end{figure}

In this figure we show constraints on the relative amplitude of $f \cdot g_{\rm a \gamma \gamma}$ by simply inserting values for the primordial magnetic field strength. It is clear that for magnetic fields near the current CMB upper bound, the \COBEF data is capable of exploring regions of parameter space not yet accessed by other experimental probes, both on the high and low mass ends. In principle, these large magnetic field scenarios should also be capable of probing the axion dark matter regime ($\Omega_{\rm a} \simeq \Omega_{\rm dm}$), though the combined interplay of $\gamma \leftrightarrow a$ conversions is beyond the scope of this work. Naturally, a \PIXIE-type experiment would allow us to probe roughly 2 orders of magnitude deeper, highlighting the value of future spectral distortion experiments \citep[e.g., as envisioned for ESA][]{Chluba2021Voyage}. 

In Fig.~\ref{fig:generic_B}, we also show the constraints one can put on the conversion probability for general (non-comoving) $B$-fields. In this case, if one knows the magnitude of the $B-$field at the conversion redshift (which we assume is set independently of $B$) constraints from axion-photon conversion can be read off from the plot. This may be useful in scenarios where magnetic fields are generated from some causal process after inflation, such as through cosmological phase transitions, or when additional amplification of a seed field occurs through the dynamo mechanism\footnote{It is well known that magnetic fields generated by dynamo mechanisms do not follow the $B(z) = B_0(1+z)^2$ scaling \citep{Banerjee2004,
Pomakov2022,Rafelski2024}.}. In order for these constraints to be applicable, the coherence length of a generic magnetic field needs to be compared against the resonance length of the axion-photon conversion, ensuring that $\ell_{\rm osc} \lesssim \ell_{\rm B}$. Finally, we restrict to the mass range in which only single resonant conversions are possible.

\section{Discussion and Conclusion}
\label{sec:level6}
\noindent In this work, we have re-analyzed constraints induced by resonant photon-to-axion conversions from the generation of CMB spectral distortions. In addition to an updated analytic treatment of both the conversion process and the distortion signature, we also perform a fully numeric treatment for the evolution of the photon occupation number using the code \texttt{CosmoTherm}. This numeric result agrees well with the analytics, providing a slight increase in sensitivity which comes from the ability to utilize the full shape of the induced distortion when comparing with data from \COBEF. Our main results are summarized in Fig.~\ref{fig:Limits_distortion} where we compare a single conversion analytic approach with different runmodes from \texttt{CosmoTherm}.

Another key improvement that we make is a more complete treatment of the photon plasma mass. This is achieved through a consistent treatment of the refractive index in the presence of not only electrons, but neutral hydrogen, as well as singly- and doubly-ionized helium. In particular, we highlight a misconception in the literature where the refractive index for molecular hydrogen (${\rm H}_2$) was most likely mistakenly used in place of the one for atomic hydrogen (HI). When considering only the HI contribution for the plasma mass, the coefficient of the $X_{\rm HI}$ term in Eq.~\eqref{eq:mgamma_def_eH_old} is roughly $70\%$ of the value which is often found in the other works on resonant conversions. For the most consistent treatment, we recommend the use of the expression in Eq.~\eqref{eq:mgamma_def_eH} which includes the helium contributions. As can be seen by the dashed line in Fig.~\ref{fig:x_gamma_crit} as well as the spikes in Fig.~\ref{fig:P_x_var_m}, helium can induce strong conversions at high frequencies. For the $\gamma \rightarrow a$ process, this turns out to be rather unimportant because this is in the Wien tail of the CMB. This could, however, give rise to some interesting phenomenology in the opposite regime where $\Omega_{\rm a} \approx \Omega_{\rm dm}$ when the $a \rightarrow \gamma$ process dominates the signal. We leave further exploration of this to future work.

Our analytic results follow from the usual Green's function approach \citep{Chluba2014, Chluba2015GreensII} for pre-recombination conversions, and a re-analysis of the \COBEF data for post-recombination conversions. This similar approach was utilized in past works \citep{mirizzi2009constraining, tashiro2013constraints} and our analytic results broadly agree with those from \citet{mirizzi2009constraining}. In addition to this consistency check, we also determined the spectral template of the axion distortion, which can be seen in Fig.~\ref{fig:A_distortion}. Note that the $M(x)$ ($\mu$) type spectral shape is a stationary solution to the photon evolution equation (for all $z \lesssim 2\times 10^6$). Unlike the dark photon case considered in \citet{Chluba2024DP}, the axion distortion does not look like $M(x)$ nor $Y(x)$, and will subsequently evolve towards $M(x)$ depending on the redshift it was sourced from. 

While only interesting for a marginal region of parameter space in the neighbourhood of $m_{\rm a} \simeq 10^{-4}$ eV, we also considered the so-called large distortion regime in which the fractional energy and entropy extraction can be large ($\epsilon_{\rho},\epsilon_{\rm N} \simeq 1$). This necessarily implies a large conversion probability ($\gamma_{\rm con} \gtrsim 1$), at which point the analytic treatment breaks down. Naively, one would have expected that for $\gamma_{\rm con} \gtrsim 0.1$, the sign of the distortion would flip from negative to positive as can be seen in Fig.~\ref{fig:Drho_rho_distortion}. The numerics, however, indicate that large distortions are accompanied by photon shock fronts in the distribution function (e.g., see Fig.~\ref{fig:Initial_large_shock}, produced by copious amounts of stimulated Compton scattering). This leads to an overall net \textit{increase} in $\epsilon_{\rm N}$, which is somewhat counterintuitive given that the axion photon conversion initially extracted photons from the CMB. It is exactly this increase in the relative entropy which enforces the negative sign of the distortion, preventing any sign flip from occurring in the overall distortion. We remark that these results assume that the Landau-Zener expression holds up to $\gamma_{\rm con} \simeq 1$. As this regime does not affect the relevant parts of our constraints, we choose not to investigate this effect further in this work.

We additionally utilized upper bounds coming from the CMB to generate constraints on $g_{\rm a \gamma \gamma}$ for a number of benchmark $B_{\rm rms}^0$ values, as shown in Fig.~\ref{fig:gagg_constraints}. The conclusion is that for large scale ($\ell_{\rm B} \simeq 1$ Mpc) magnetic field strengths near this upper limit, it is possible for CMB distortions to explore regions of parameter space untouched by other phenomenology. The constraints fall off roughly linearly with the field strength, however, and for $B_{\rm rms}^0 \lesssim 10^{-2}$ nG we find that data from the \COBEF satellite becomes a suboptimal way to search for ALPs. Forecasts for a \PIXIE-type satellite improve the constraining power by roughly 2 orders of magnitude, implying that primordial $B$-fields at the $10^{-4} \, $nG level today could generate axion-induced spectral distortions.

Despite the \COBEF dataset being nearly three decades old, CMB spectral distortions are still capable of providing leading constraints on models of axion-photon conversions. This longevity of relevance underscores the value of precision measurements of the CMB frequency spectrum, both for gaining a deeper understanding of the standard cosmological model, and for the constraining power offered to dozens of BSM scenarios.

\section*{Acknowledgements}
\noindent We would like to thank Levon Pogosian for insightful discussions on primordial magnetic fields. BC would like to acknowledge support from both an NSERC Banting fellowship, as well as the Simons Foundation (Grant Number 929255).

\bibliographystyle{apsrev4-1}
\bibliography{Lit-axion.bib, Lit.bib}

\end{document}